\documentclass[%
 aip,
 amsmath,amssymb,
 reprint,%
]{revtex4-1}

\usepackage{graphicx}
\usepackage{dcolumn}
\usepackage{bm}

\usepackage[utf8]{inputenc}
\usepackage[T1]{fontenc}
\usepackage{mathptmx}
\usepackage{xspace, microtype}
\usepackage{siunitx, makecell}
\usepackage{soul}
\usepackage{hyperref}

\usepackage{filecontents}
\usepackage[dvipsnames]{xcolor}
\usepackage{cleveref}
\usepackage{afterpage}

\newcommand\myshade{85}
\colorlet{mylinkcolor}{BrickRed}
\colorlet{mycitecolor}{NavyBlue}
\colorlet{myurlcolor}{Aquamarine}

\hypersetup{
  linkcolor  = mylinkcolor!\myshade!black,
  citecolor  = mycitecolor!\myshade!black,
  urlcolor   = myurlcolor!\myshade!black,
  colorlinks = true,
}
\newcommand\subref[2]{\hyperref[#1]{\ref*{#1}#2}}

\newcommand{\phii}{\ensuremath{\Phi}\xspace}
\newcommand{\Phii}{\phii}
\DeclareMathOperator{\var}{var}
\newcommand{\PhiR}{\ensuremath{\phii^{\mathrm{R}}}\xspace}
\newcommand{\phiid}{\ensuremath{\Phi\mathrm{ID}}\xspace}

\makeatletter
\DeclareFontEncoding{LS1}{}{}
\DeclareFontEncoding{LS2}{}{\noaccents@}
\DeclareFontSubstitution{LS1}{stix}{m}{n}
\DeclareFontSubstitution{LS2}{stix}{m}{n}
\makeatother
\DeclareMathAlphabet\mathscr{LS1}{stixscr}{m}{n}
\SetMathAlphabet\mathscr{bold}{LS1}{stixscr}{b}{n}
\DeclareMathAlphabet\mathcal{LS2}{stixcal}{m}{n}
\SetMathAlphabet\mathcal{bold}{LS2}{stixcal}{b}{n}

\definecolor{BrewerRed}{RGB}{228,26,28}
\definecolor{BrewerBlue}{RGB}{55,126,184}

\usepackage{ctable}
\graphicspath{ {./figs/} }

\begin{document}

\preprint{AIP/123-QED}

\title[Integrated Information as a Common Signature of Dynamical and Information-processing Complexity]{Integrated Information as a Common Signature of Dynamical and Information-Processing Complexity}

\author{Pedro A.M. Mediano}
\thanks{P.M. and F.R. contributed equally to this work.\\E-mail: pam83@cam.ac.uk, f.rosas@imperial.ac.uk}
\affiliation{Department of Psychology, University of Cambridge, Cambridge CB2 3EB}

\author{Fernando E. Rosas}
\thanks{P.M. and F.R. contributed equally to this work.\\E-mail: pam83@cam.ac.uk, f.rosas@imperial.ac.uk}
\affiliation{Centre for Psychedelic Research, Department of Brain Science, Imperial College London, London SW7 2DD}
\affiliation{Data Science Institute, Imperial College London, London SW7 2AZ}
\affiliation{Centre for Complexity Science, Imperial College London, London SW7 2AZ}

\author{Juan Carlos Farah}
\affiliation{School of Engineering, \'Ecole Polytechnique F\'ed\'erale de Lausanne, CH-1015 Lausanne}

\author{Murray Shanahan}
\affiliation{Department of Computing, Imperial College London, London SW7 2RH}

\author{Daniel Bor}
\affiliation{Department of Psychology, University of Cambridge, Cambridge CB2 3EB}

\author{Adam B. Barrett}
\affiliation{Sackler Center for Consciousness Science, Department of Informatics, University of Sussex, Brighton BN1 9RH}
\affiliation{The Data Intensive Science Centre, Department of Physics and Astronomy, University of Sussex, Brighton BN1 9QH}

\begin{abstract}

The apparent dichotomy between information-processing and dynamical approaches
to complexity science forces researchers to choose between two diverging sets
of tools and explanations, creating conflict and often hindering scientific
progress. Nonetheless, given the shared theoretical goals between both
approaches, it is reasonable to conjecture the existence of underlying common
signatures that capture interesting behaviour in both dynamical and
information-processing systems. Here we argue that a pragmatic use of
Integrated Information Theory (IIT), originally conceived in theoretical
neuroscience, can provide a potential unifying framework to study complexity in
general multivariate systems.
Furthermore, by leveraging metrics put forward by the integrated information decomposition 
($\Phi$ID) framework, our results reveal that integrated information can effectively
capture surprisingly heterogeneous signatures of complexity --- including
metastability and criticality in networks of coupled oscillators as well as
distributed computation and emergent stable particles in cellular automata --- without 
relying on idiosyncratic, ad-hoc criteria. These results show how an
agnostic use of IIT can provide important steps towards bridging the gap
between informational and dynamical approaches to complex systems.

\end{abstract}

\maketitle

\begin{quotation}

Originally conceived within theoretical neuroscience, Integrated Information
Theory (IIT) has been rarely used in other fields --- such as complex systems
or non-linear dynamics --- despite the great value it has to offer. In this
article we inspect the basics of IIT, dissociating it from its contentious
claims about the nature of consciousness. \mbox{Relieved} of this philosophical
burden, IIT presents itself as an appealing formal framework to study
complexity in biological or artificial systems, applicable in a wide range of
domains. To illustrate this, we present an exploration of integrated
information in complex systems and relate it to other notions of complexity
commonly used in systems such as coupled oscillators and cellular automata.
Through these applications, we advocate for IIT as a valuable framework capable
of revealing common threads between diverging branches of complexity science.

\end{quotation}

\section{Introduction}

Most theories about complexity are rooted in either information theory or
dynamical systems perspectives --- two disciplines with very different aims and
toolkits. The former, built after the work of Turing and Shannon, focuses
mainly on discrete systems and considers complexity in terms of information
processing, universal computation, distributed computation, and coherent
emergent structures.\cite{Wolfram2002} The latter, following the tradition
started by Poincar\'e and Birkhoff, focuses on continuous systems and studies
their behaviour using attractors, phase transitions, chaos, and
metastability.\cite{Pikovsky2001}

This methodological divide has contraposed how various communities of
researchers think about complex systems, even to the extent of triggering some
longstanding disagreements. This is particularly evident in the field of
cognitive neuroscience, where proponents of computational approaches claim that
the brain works similarly to a Turing
machine,\cite{fodor1975language,pylyshyn1986computation,rescorla2017ockham,milkowski2013explaining}
while opponents believe that cognitive processes are essentially continuous and
rate-dependent.\cite{van1995might,van1998dynamical,smith2005cognition,schoner2008dynamical}
A related debate has taken place in the artificial intelligence community
between symbolic and connectionist paradigms for the design of intelligent
systems.\cite{minsky1991logical} Modern stances on these problems have promoted
hybrid approaches,\cite{Garnelo2019} bypassing ontological arguments towards
epistemological perspectives where both information and dynamics represent
equally valid methods for enquiry.\cite{beer2015information}

Interestingly, bridging the gap between the information-processing and
dynamical systems literature has proven scientifically fruitful. Examples of
this are Wolfram's categorisation of cellular automata in terms of their
attractors, which provided insights into the possible types of distributed
computation enabled by these systems according to dynamical properties of their
trajectories,\cite{Wolfram1984} and Langton's intuition that computation takes
place in a phase transition ``at the edge of chaos.''\cite{Langton1990} This
rich point of contact, in turn, suggests that what informational and dynamical
approaches deem as interesting might have a common source, beyond the apparent
dissimilarities introduced by heterogeneous tools and disciplinary boundaries.

In this article we propose \emph{Integrated Information Theory}
(IIT),\cite{Tononi2003,Balduzzi2008,Oizumi2014} together with its recent
extension \emph{Integrated Information Decomposition}
($\Phi$ID),\cite{Mediano2019a} as a conceptual framework that can help bridge
the gap between information-processing and dynamical systems approaches. At its
inception, IIT was a theoretical effort that attempts to explain the origins
and fundamental nature of consciousness.\cite{Tononi2015,Tononi2016} The
boldness of IIT's claims have not gone unnoticed, and they have caused heated
debate in the neuroscience community.\cite{Barrett2019,Cerullo2015,Mindt2017}
Unfortunately, its audacious claims about consciousness have kept many
scientists away from IIT, thereby preventing some of its valuable theoretical
insights from reaching other areas of knowledge.

We advocate for the adoption of a \emph{pragmatic} IIT, that can be used to
analyse and understand complex systems without the philosophical burden of its
origins in consciousness science. Consequently, the goal of this paper is to
dissociate IIT's claims as a theory of consciousness from its formal
contributions, and put the latter to use in the context of complexity science.
For this purpose, we demonstrate that integrated information --- as calculated
in $\Phi$ID --- peaks sharply for oscillatory dynamical systems that exhibit
criticality and metastability, and also characterises cellular automata that
display distributed computation via persistent emergent structures. These
findings illustrate the remarkable flexibility of integrated information
measures in discovering features of interest in a wide range of scenarios,
without relying on domain-specific considerations. Overall, this work reveals
how a grounded, demystified interpretation of IIT can allow us to identify
features that are transversal across complex information-processing and
dynamical systems.

The rest of this article is structured as follows. Section~\ref{sec:core}
presents the core formal ideas of IIT and $\Phi$ID in a simple manner, and puts
forward a sound and demystified way of interpreting its key quantities. Our
main results are presented in Sections~\ref{sec:osc} and \ref{sec:automata}:
the former shows that integrated information can capture metastability and
criticality in dynamical systems, and the latter that integrated information is
a distinctive feature of distributed computation. Finally,
Section~\ref{sec:discussion} discusses the significance of these results, and
summarises our main conclusions.

\section{A pragmatist's IIT}
\label{sec:core}

IIT constitutes one of the first attempts to formalise what makes a system
``more than the sum of its parts,'' building on intuitive notions of synergy
and emergence that have been at the core of complexity science since its
origins.\cite{anderson1972more,Waldrop1993} IIT proposes \emph{integrated
information} for that role, informally defining it as information that is
contained in the interactions between the parts of a system and not within the
parts themselves. The core element of IIT is \phii, a scalar measure that
accounts for the amount of integrated information present in a given system.

While faithful to its original aims, throughout its life IIT has undergone
multiple revisions. Of all of them, we will focus on the theory as originally
introduced by Balduzzi \& Tononi in 2008.\cite{Balduzzi2008} While more recent
accounts of the theory exist,\cite{Oizumi2014} these place a much stronger
emphasis on its goals as a theory of consciousness, at the expense of a
departure from standard information-theoretic practice and more convoluted
algorithms --- which have hindered its reach and made the theory applicable
only in small discrete systems.

\subsection{The maths behind $\Phi$}
\label{sec:math_behind_phi}

\begin{figure*}[t]
  \centering
  \includegraphics{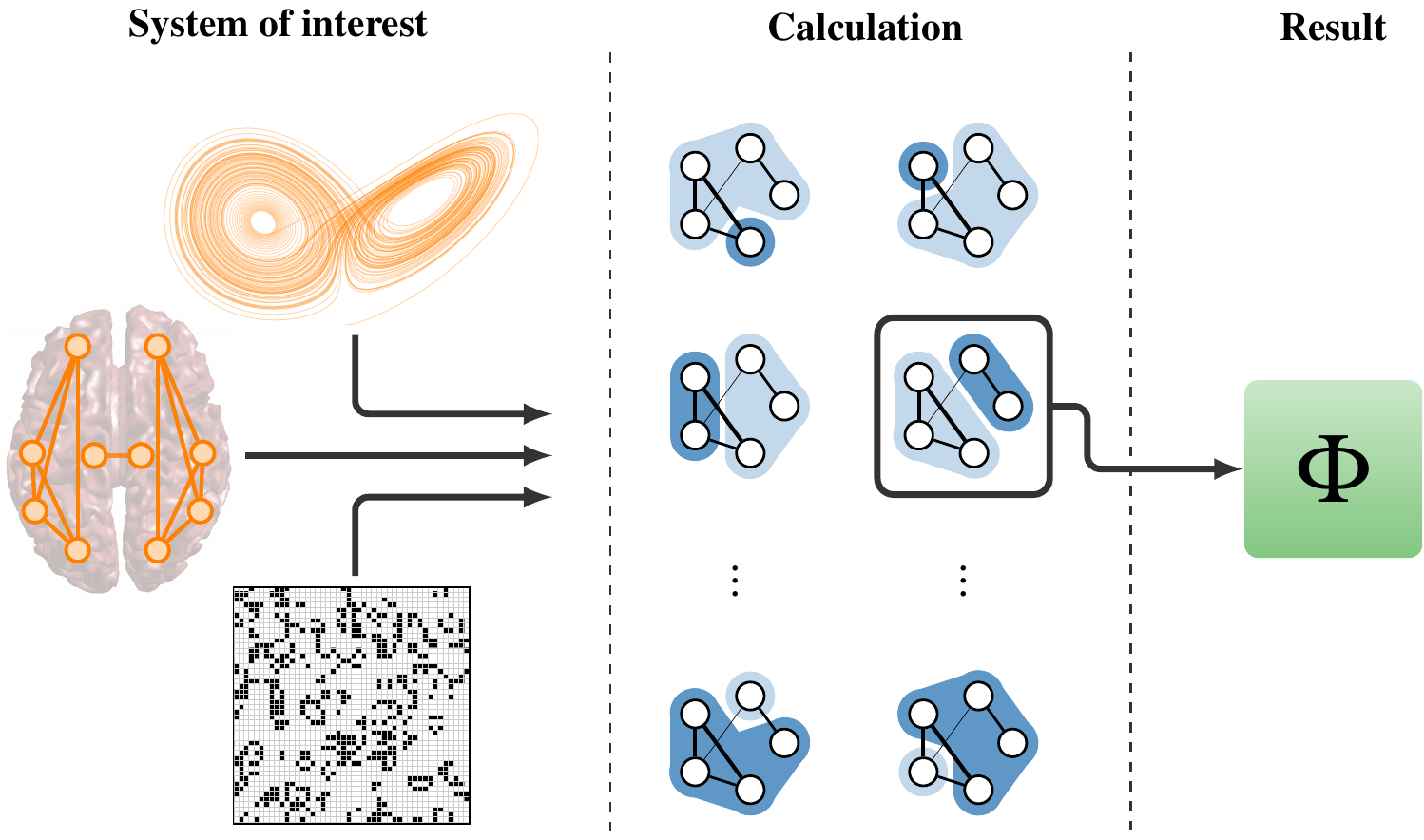}

  \caption{\textbf{Graphical illustration of an integrated information analysis
    in complex systems}. Integrated Information Theory (IIT) can be used to
    analyse a wide range of complex systems, from dynamical systems to cellular
    automata and empirical data (\emph{left}). Due to the generality of
    information-theoretic measures, it can be applied to either real- or
    discrete-valued data. After a suitable statistical model has been
    estimated, the system is partitioned following Eq.~\eqref{eq:phi}, and
    effective information $\varphi$ is computed for each partition
    (\emph{middle}). Finally, the partition with the ``cruelest cut'' (more
  formally, the minimum information partition) is selected, and the final value
of integrated information \phii is computed (\emph{right}).}

  \label{fig:iit_diagram}
\end{figure*}

This section provides a succinct description of the mathematical formulae
behind IIT 2.0,\cite{Balduzzi2008} following Barrett \&
Seth's\cite{Barrett2011} concept of \emph{empirical} integrated information.
The overall analysis procedure is represented schematically in
Figure~\ref{fig:iit_diagram}.

The building block of integrated information is a measure of \emph{effective
information} (i.e. excess of predictive information) typically denoted as
$\varphi$.\cite{Balduzzi2008} Effective information quantifies how much better
a system $X$ is at predicting its own future after a time $\tau$ when it is
considered as a whole compared to when it is considered as the sum of two
subsystems $M^{1}$ and $M^{2}$ (so that $X=(M^{1},M^{2})$). In other words,
$\varphi$ evaluates how much predictive information is generated by the system
over and above the predictive information generated by the two subsystems
alone. For a given bipartition $\mathcal{B} = \{M^1,M^2\}$, the effective
information of the system $X$ beyond $\mathcal{B}$ is calculated as
\begin{align}
  \varphi[X; \tau, \mathcal{B}] = I(X_{t-\tau} ;  X_t) - \displaystyle\sum_{k=1}^2 I(M_{t-\tau}^k ;
  M_t^k) ~ ,
  \label{eq:ei}
\end{align}
\noindent where $I$ is Shannon's mutual information. We refer to $\tau$ as the
\emph{integration timescale}.

The core idea behind the computation of $\Phi$ is to (i) exhaustively search
all possible partitions of the system, (ii) calculate $\varphi$ for each of
them, and (iii) select the partition with lowest $\varphi$ (under some
considerations, see below), termed the \emph{Minimum Information Bipartition}
(MIB). Then, the integrated information of the system is defined as the
effective information beyond its MIB. Concretely, the integrated information
$\Phi$ associated with the system $X$ over the integration scale $\tau$ is
given by
\vspace{2pt}
\begin{subequations}
\begin{gather}
\Phi[X;\tau] = \varphi[X; \tau, \mathcal{B}^{\mathrm{MIB}}]~,\\
\mathcal{B}^{\mathrm{MIB}} = \arg_{\mathcal{B}}\min \frac{\varphi[X; \tau,
\mathcal{B}]}{K(\mathcal{B})}, \\
  K(\mathcal{B}) = \min \left\{ H(M^1), H(M^2) \right\} ,
\end{gather}
\label{eq:phi}
\end{subequations}
\noindent where $K$ is a normalisation factor introduced to avoid biasing
$\Phi$ to excessively unbalanced bipartitions. Defined this way, $\Phi$ can be
understood as the minimum information loss incurred by considering the whole
system as two separate subsystems.

An important drawback of \phii is that it can take negative values, which
hinders its interpretation as a measure of system-wide
integration.\cite{Mediano2019} Recently, Mediano~\emph{et
al.}\cite{Mediano2019a} showed that \phii quantifies not only information
shared across and transferred between the parts, but also includes a negative
component measuring \emph{redundancy} -- i.e. when the parts contain \emph{the
same} predictive information. Therefore, \phii measures a balance between
information transfer and redundancy, such that $\phii < 0$ when the system is
redundancy-dominated.

A principled way to address this limitation is to refine $\Phi$ by
disentangling the different information phenomena that drive it. This can be
achieved via the $\Phi$ID framework,\cite{Mediano2019a} which provides
principled tools to study the different information phenomena that can take
place in a multivariate dynamical system. Using $\Phi$ID, one can define a
revised version of $\varphi$, denoted as
$\varphi^{\mathrm{R}}$,\cite{Mediano2019a} which removes the redundancy
component in $\Phi$. A simple and effective way to compute
$\varphi^{\mathrm{R}}$ is via the Minimum Mutual Information
(MMI)\cite{Barrett2015} redundancy function, which allows us to compute
$\varphi^{\mathrm{R}}$ by 'adding back' the redundancy:
\begin{align}
  \varphi^{\mathrm{R}}[X; \tau, \mathcal{B}] = \varphi[X; \tau, \mathcal{B}] + \min_{i,j}I(M^i_{t-\tau}; M^j_t)~.
  \label{eq:phiR}
\end{align}
Using $\varphi^{\mathrm{R}}$, we can define \PhiR analogously through
Eqs.~\eqref{eq:phi}.

Note that this revised measure of integrated information not only has better
theoretical properties than the traditional $\varphi$, as discussed by Mediano
\textit{et al},~\cite{Mediano2019a}, but also has been observed to be superior
in practical neuroimaging analyses.\cite{luppi2020synergistic}

\subsection{Interpretation of $\Phii$}
\label{sec:interpretation}

Conceptually, there are several ways to interpret \phii, which highlight
different aspects of the systems under study. One interpretation ---
particularly relevant for complexity science~--- is based on the theory of
\textbf{information dynamics}, which decomposes information processing in
complex systems in terms of storage, transfer, and
modification.\cite{Lizier2010,Lizier2007,Lizier2012,Lizier2010c} From this
perspective, and based on earlier results,\cite{Mediano2019a} \PhiR can be seen
as capturing a combination of information modification across multiple parts of
the system, information transfer from one part to another, and storage in
coherent structures that span across more than one system variable.

Alternatively, a more quantitative and mathematically rigorous way of
interpreting \PhiR is in terms of \textbf{prediction bounds}. The conditional
entropy (a reciprocal of the mutual information\cite{james2011anatomy})
provides an upper bound on the optimal prediction
performance,\cite{hellman1970probability,feder1994relations} such that a system
with low conditional entropy can be predicted accurately. Therefore, mutual
information acts as a bound too: the higher the mutual information between two
variables, the better one can be predicted from the other. Thus, \PhiR measures
to what extent the full state of the system enables better predictions than the
states of the parts separately.

Note that other interpretations of \phii-related quantites exist (most notably
through information geometry and statistical
inference\cite{cohen2020general,Langer2020}), although they do not apply as
cleanly to Eqs.~\eqref{eq:ei} and \eqref{eq:phiR}.

\section{Integrated information, metastability, and phase transitions}
\label{sec:osc}

\begin{figure*}[t]
  \centering
  \includegraphics{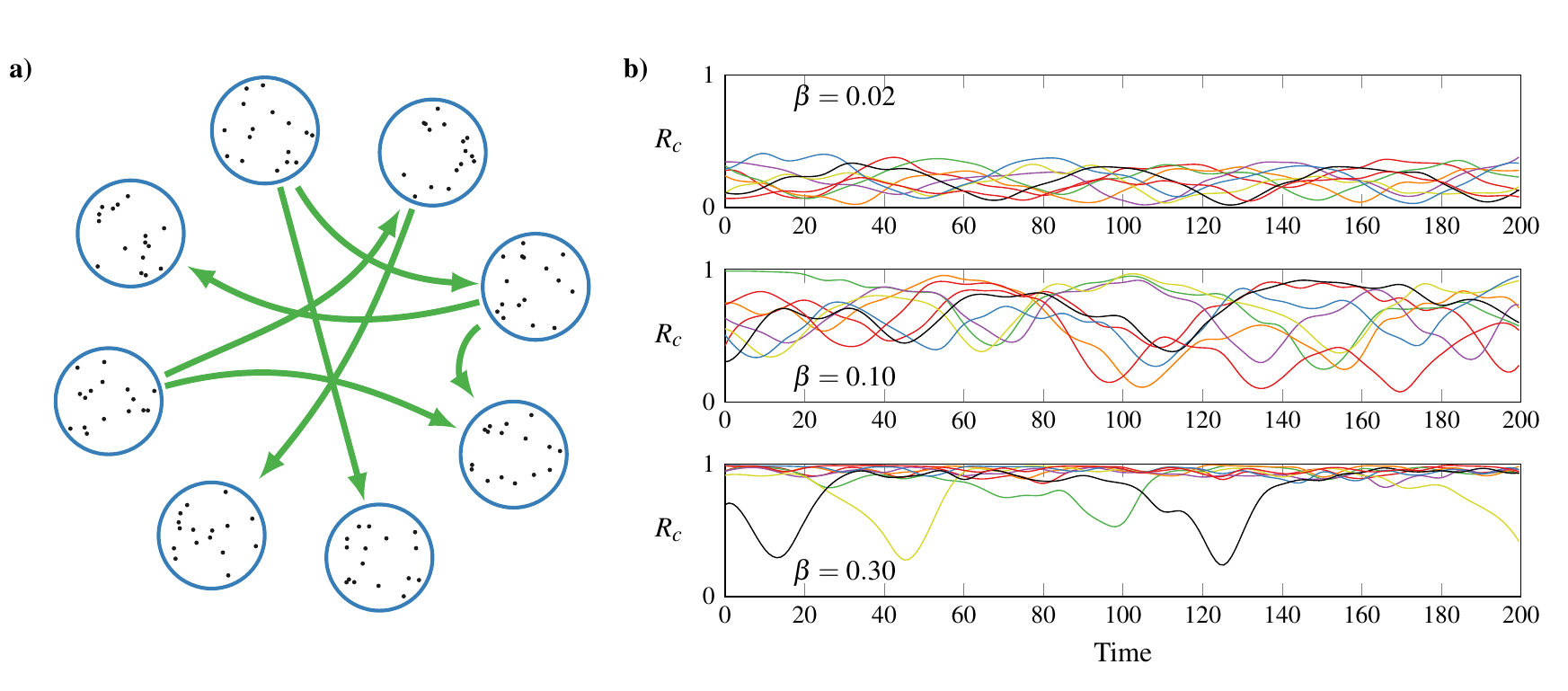}

  \caption{\textbf{A community-structured network of coupled oscillators
    exhibits metastable chimera-like behaviour}. \textbf{(a)} We study a
    network formed by 8 communities of 32 Kuramoto oscillators each, with
    strong intra-community coupling and weak inter-community coupling, and
    dynamics parameterised by a phase lag $\beta$. \textbf{(b)} Time series of
    synchrony values for each community ($c = 1, \ldots, 8$) show that the
    behaviour of the system changes drastically with $\beta$. For low $\beta$
    all communities remain desynchronised; for high $\beta$ all communities
  synchronise; and it is for intermediate $\beta$ that the system behaves as a
metastable chimera, with different communities entering intermittent periods of
internal synchrony.}

  \label{fig:kuramoto}
\end{figure*}

This section explores the usage of integrated information to study dynamical
systems, exploring the relationship between \Phii, metastability, and phase
transitions. For this, we focus on systems of coupled oscillators, which are
ubiquitous in both natural and engineered environments, making them of
considerable scientific interest.\cite{Pikovsky2001} Typical studies of
oscillatory systems --- like the classic work of Kuramoto \cite{Kuramoto1984}
--- examine the conditions under which the system stabilises on states of
either full synchronisation or desynchronisation; although these two extremes
are by no means representative of all real-world synchronisation phenomena.
Many systems of interest, including the human brain, exhibit synchronous
rhythmic activity on multiple spatial and temporal scales but never settle into
a stable state, entering so-called chimera-like states\cite{Panaggio2015} of
high partial synchronisation only temporarily. A system of coupled oscillators
that continually moves between highly heterogeneous states of synchronisation
is said to be \emph{metastable}.\cite{Shanahan2010}

In 2010, Shanahan\cite{Shanahan2010} showed that a modular network of
phase-lagged Kuramoto oscillators can exhibit metastable chimera states.
Variants of this model have since been used to replicate the statistics of the
brain under a variety of conditions, including wakeful rest\cite{Cabral2011}
and anaesthesia.\cite{Schartner2015} In the following, we study these
metastable oscillatory systems through the lens of integrated information
theory.

\subsection{Model and measures}

We examine a community-structured network of coupled Kuramoto oscillators
(shown in Fig.~\ref{fig:kuramoto}), building on the work of
Shanahan.\cite{Shanahan2010} The network is composed of $N$ communities of $m$
oscillators each, with every oscillator being coupled to all other oscillators
in its community, and to each oscillator in the rest of the network with
probability $q$. The state of the $i$-th oscillator is determined by its phase
$\theta_i$, the evolution of which is governed by
\begin{align}
  \frac{d \theta_i}{d t} = \omega + \frac{1}{\kappa} \displaystyle\sum_j K_{ij} \sin \left( \theta_j - \theta_i - \alpha \right) ~ ,
  \label{eq:kur}
\end{align}
\noindent where $\omega$ is the natural frequency of the oscillators, $\alpha$
is a global \emph{phase lag}, $K_{ij}$ are the connectivity coefficients, and
$\kappa$ is a normalisation constant. To reflect the community structure, the
coupling between two oscillators $i,j$ is defined as
\begin{equation}
    K_{i,j} =
    \begin{cases}
    a & \text{if $i$ and $j$ are in the same community, or}\\
    b & \text{otherwise,}
    \end{cases}
\end{equation}
with $a>b$. The system is tuned by modifying the value of the phase lag,
parametrised by $\beta = \pi/2 - \alpha$. We note that the system is fully
deterministic, i.e. there is no noise injected in the dynamical equations.

To assess the dynamical properties of the oscillators we consider their
\emph{instantaneous synchronisation} $R$ and \emph{metastability index}
$\lambda$. The instantaneous synchronisation of community $c\in\{1,\dots,N\}$
at time $t$ quantifies the dispersion in $\theta$-space of the corresponding
oscillators:
\begin{equation}
  R_c(t) = \left| \big\langle e^{i \theta_j(t) } \big\rangle_{j \in c} \right| ~ .
  \label{eq:sync}
\end{equation}
Building on this notion, the metastability of each community is defined as the
variance of synchrony over time, and the overall metastability is its average
across communities:
\begin{subequations}
\begin{gather}
  \lambda_c = \var_t R_c(t) \label{eq:lambdac} \\
  \lambda = \langle \lambda_c \rangle_c ~ .
\end{gather}
\label{eq:lambda}
\end{subequations}
Communities that are either hypersynchronised or completely desynchronised are
both characterised by small values of $\lambda_c$, whereas only communities
whose elements fluctuate in and out of synchrony have a high $\lambda_c$. Put
simply, a system of oscillators exhibits metastability to the extent that its
elements fluctuate between states of high and low synchrony. In addition to
metastability, we also consider the \emph{global synchrony} of a network
defined as the spatio-temporal average of the instantaneous synchrony,
\begin{equation}
  \xi = \left| \big\langle R_c(t) \big\rangle_{t,c} \right| ~ .
\end{equation}

For tractability, we calculate $\PhiR$ with respect to the \emph{coalition
configuration} of the system, defined for each community $c$ and time $t$ as
\begin{equation*} X_t^c = \begin{cases} 1 & \text{if } R_c(t) > \gamma\\ 0 &
\text{otherwise,} \end{cases} \end{equation*}
where $\gamma$ is the coalition threshold. This representation provides $N$
interdependent binary time series $X_t:=(X_t^1,\dots,X_t^N)$, which indicates
the set of communities that are more internally synchronised. 
The Shannon entropy of $X_t$ is referred to as the \emph{coalition entropy} $H_c$,
and quantifies the diversity of synchronisation patterns across communities.

\subsection{Results}

We simulated a network composed of $N=8$ communities of $m=32$ oscillators
each. The probability of connections accross communities was set to $q=1/8$,
with connection strengths of $a=0.6$ within communities and $b=0.4$ across. The
natural frequency used was $\omega = 1$, and the normalisation constant
$\kappa=64$. We ran 1500 simulations with values of $\beta$ distributed
uniformly at random in the range $[0, 2\pi)$ using a
4\textsuperscript{th}-order Runge-Kutta algorithm, using a step size of 0.05
for numerical integration. Each simulation was run for \num{5e6} timesteps,
discarding the first \num{e4} to avoid transient effects and applying a
thinning factor of 5. For the results presented here we used $\gamma = 0.8$,
and we confirmed that results were qualitatively stable for a wide range of
threshold values. All information-theoretic measures are reported in bits.

\subsubsection{Metastability and $\PhiR$ at the phase transition}

\begin{figure}[t]
\centering
\includegraphics{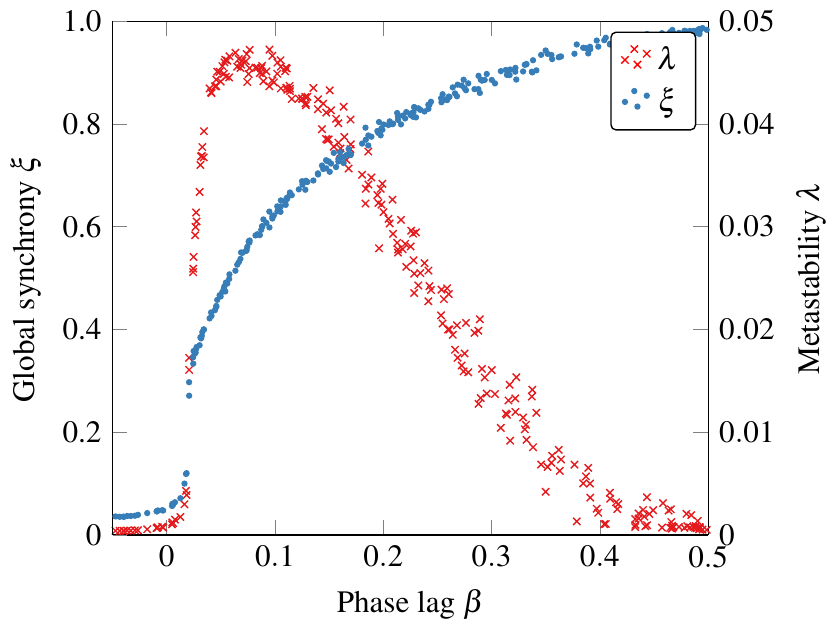}

\caption{\textbf{Phase transition between desynchronisation and
  hypersynchronisation in coupled oscillators}. Global synchrony $\xi$ and
  metastability $\lambda$ for different phase lags $\beta$ in the critical
  region. Rapid increase of metastability marks the onset of the phase
transition.}

\label{fig:sync}
\end{figure}

We first study the system from a purely dynamical perspective, and, replicating
previous findings,\cite{Shanahan2010} find two well differentiated dynamical
regimes: one of hypersynchronisation and one of complete desynchronisation,
with strong metastability appearing in the narrow transition bands between them
(Fig.~\ref{fig:sync}). Interestingly, it is in this transition region where the
oscillators operate in a critical regime poised between order and disorder, and
where complex phenomena appear. As the system moves from desynchronisation to
full synchronisation there is a sharp increase in metastability, followed by a
smoother decrease as the system becomes hypersynchronised.

\begin{figure}[t]
  \centering
  \includegraphics{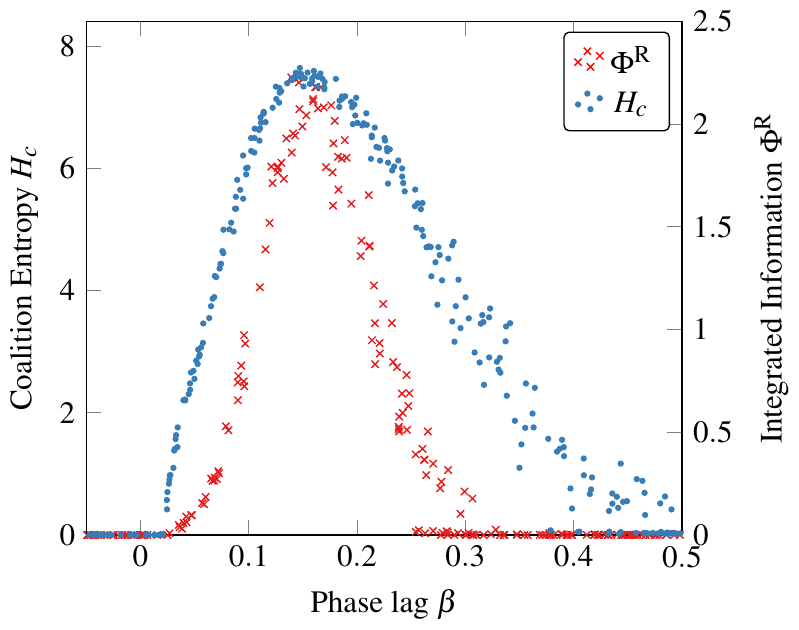}

  \caption{\textbf{Integrated information $\PhiR$ peaks in the phase transition
  of coupled oscillators}. Within the broad region between order and disorder
in which $H_c$ rises there is a narrower band in which complex spatio-temporal
patterns generate high $\PhiR$. Calculations use $\tau=100$.}

\label{fig:phi}
\end{figure}

Importantly, $\PhiR$ was found to exhibit a similar behaviour to $\lambda$: it
is zero for desynchronised systems, peaks in the transition region, and shrinks
again in the fully ordered regimes (Fig.~\ref{fig:phi}). This shows that
networks in metastable regimes are the only ones that exhibit integrated
information. When comparing $\PhiR$ with the coalition entropy, results show
that both peak precisely at the same point, although the peak in $\PhiR$ is
much narrower than the peaks in $\lambda$ and $H_c$. Hence, while some values
of $\beta$ do give rise to non-trivial dynamics, it is only at the centre of
the critical region that these dynamics give rise to integrated information.

These results imply that $\PhiR$ is sensitive to more subtle dynamic patterns
than the other measures considered, and is in that sense more discriminating.
In effect, a certain degree of internal variability is necessary to establish
integrated information, but not all configurations with high internal
variability lead to a high $\PhiR$. Also, $\PhiR$ accounts for spatial
\emph{and} temporal patterns in a way that the other metrics do
not.\footnote{This can be verified by performing a random time-shuffle on the
time series, which leaves $\lambda$ and $H_c$ unaltered as they do not
explicitly depend on time correlations, but makes $\PhiR$ shrink to zero.}

\subsubsection{Integrated information at multiple timescales}
\label{sec:phi_timescales}

\begin{figure}[t]
  \centering
  \includegraphics{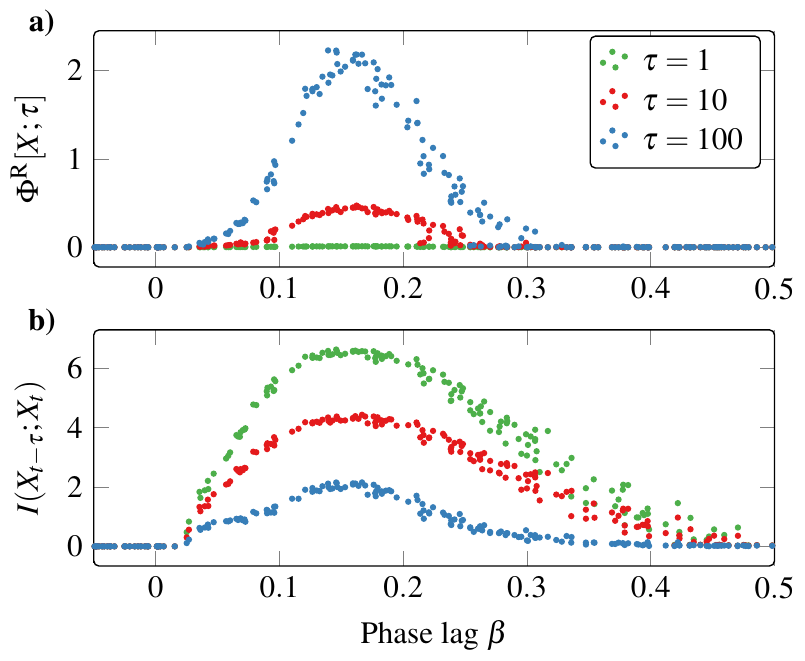}

  \caption{\textbf{Redundancy dominates at shorter timescales, synergy at
    longer ones}. \textbf{(a)} Integrated information $\PhiR$ grows with
    increasing integration timescale $\tau$, indicating a shift from
    redundancy- to synergy-dominated dynamics. \textbf{(b)} Time-delayed mutual
  information $I(X_{t - \tau}; X_t)$ decreases with higher $\tau$, indicating
an overall loss of predictability for longer time horizons.}

  \label{fig:phiManyTau}
\end{figure}

As a further analysis, we can investigate the behaviour of $\PhiR$ at multiple
timescales by varying the integration timescale parameter $\tau$ (see
Fig.~\ref{fig:kuramoto} for a visual guide of different $\tau$ values).
Fig.~\ref{fig:phiManyTau} shows $\PhiR$ for several values of $\tau$, and
compares it with standard time-delayed mutual information (TDMI) $I(X_{t -
\tau}; X_t)$. Note that this analysis cannot be carried out with $H_c$, or
other measures of complexity that are not sensitive to temporal order --- i.e.
that are functions of $p(X_t)$, and not $p(X_t | X_{t-\tau})$.

Results show that $\PhiR$ and TDMI exhibit opposite trends with respect to
changes in $\tau$: TDMI decreases with higher $\tau$, while $\PhiR$ increases.
At short timescales the system is highly predictable --- thus the high TDMI ---
but this short-term evolution does not imply much system-wide interaction ---
thus the low $\PhiR$.
Together, the high TDMI and low $\PhiR$ suggest that at short timescales the
system is redundancy-dominated: the system contains information about its
future, but this information can be obtained from the parts separately.
Conversely, for prediction at longer timescales, TDMI decreases but \PhiR
increases, indicating that while the system is overall less predictable, this
prediction is enabled by the information contained in the interaction between
the parts.

\subsubsection{Robustness of $\PhiR$ against measurement noise}

\begin{figure}[t]
  \centering
  \includegraphics{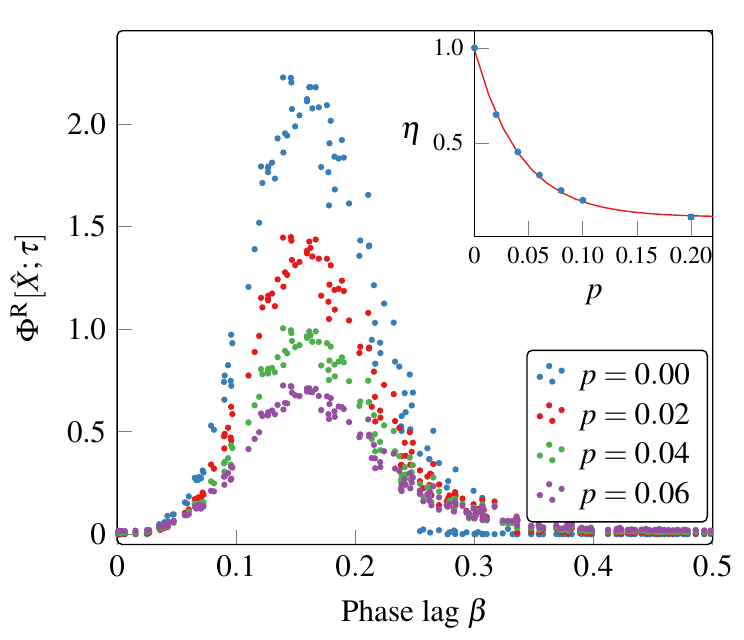}

  \caption{\textbf{Integrated information is highly sensitive to measurement
    noise}. Integrated information $\PhiR$ for different levels of measurement
    noise $p$. Inset shows the mean and variance of the ratio $\eta$ between
    $\PhiR$ of the corrupted and the original time series
  (\textcolor{BrewerBlue}{blue}), and an exponential fit $\eta =
\exp(-p/\ell)$, with $\ell \approx 0.04$ (\textcolor{BrewerRed}{red}).}

  \label{fig:phiNoise}
\end{figure}

Finally, we study the impact of measurement noise on $\PhiR$, wherein the
system runs unchanged but our recording of it is imperfect. For this, we run
the (deterministic) simulation as before and generate the sequence of coalition
configurations, and then emulate the effect of uncorrelated measurement noise
by flipping each bit in the time series with probability $p$, yielding a
corrupted time series $\hat{X}_t$. Finally, $\PhiR$ is recalculated on the
corrupted time series (Fig.~\ref{fig:phiNoise}). To quantify the impact of
noise, we studied the ratio between the corrupted and the original time series,
\begin{equation} 
  \eta = \frac{\PhiR[\hat{X}; \tau]}{\PhiR[X; \tau]} ~ .
  \label{eq:eta} 
\end{equation}
To avoid instabilities as $\PhiR[X; \tau] \approx 0$, we calculate $\eta$ only
in the region within \SI{0.5}{\radian} of the centre of the peak shown in
Fig.~\ref{fig:phi}, where $\PhiR[X; \tau]$ is large. The inset of
Fig.~\ref{fig:phiNoise} shows the mean and standard deviation of $\eta$ at
different noise levels.

Results show that $\PhiR$ decays exponentially with $p$
(Fig.\ref{fig:phiNoise}, upper pannel), reflecting a gradual loss of the
precise spatio-temporal patterns that are characteristic of the system. In
particular, $\PhiR$ was found to be highly sensitive to noise and to undergo a
rapid decline, as a measurement noise of 5\% can wipe out 70\% of the observed
integrated information of the system. While the distortion has a stronger
effect on time series with greater $\PhiR$, it preserves the dominant peak for
all values of $p$.

\begin{figure*}[t]
  \centering
  \includegraphics{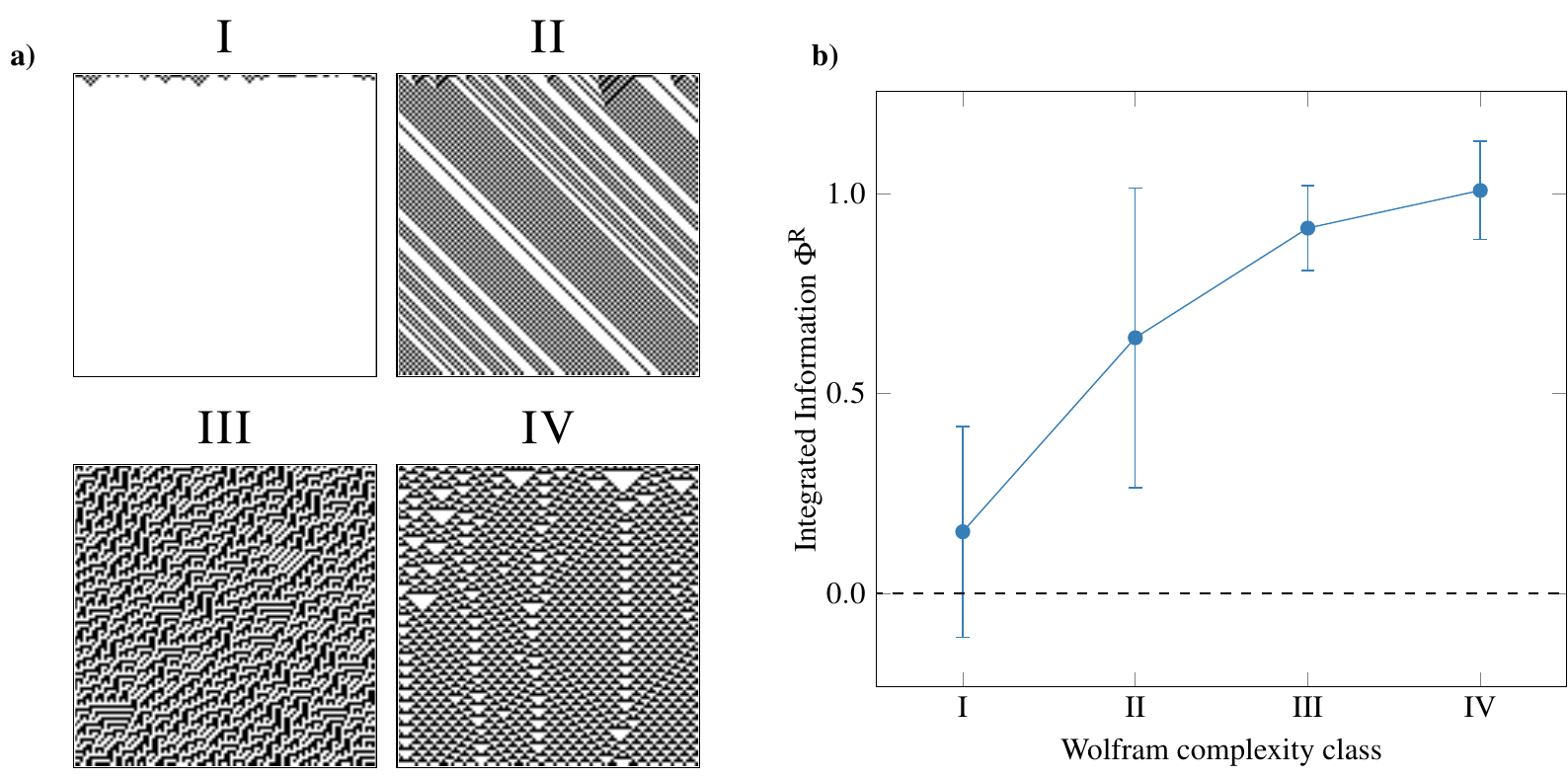}

  \caption{\textbf{Integrated information grows monotonically with Wolfram
    class number}. \textbf{(a)} Examples of each complexity class (ECA rules
    32, 56, 75, and 54, respectively), showing noticeable differences in
    behaviour.  Notice the presence of localised particles in the class
    \textsc{IV} rule.  \textbf{(b)} Correspondingly, \PhiR is highest for the
    more complex classes \textsc{IV} and \textsc{III}, and lower (and often
    negative) for the simpler behaviours in classes \textsc{I} and \textsc{II}
  (error bars correspond to standard deviation across rules; each rule was
simulated multiple times to obtain an accurate estimation of \PhiR).}

\label{fig:phiwolfram}
\end{figure*}

Overall, in this section we have shown that a network of Kuramoto oscillators
presents a sharp, clear peak of integrated information around its phase
transition that coincides with a strong increase in metastability. Furthermore,
we have found that \PhiR is informative, as it can reveal information about
timescales of interaction between system components. Finally, \PhiR was also
found to be sensitive, as it vanishes quickly if the specific spatio-temporal
patterns of the system under study are disrupted. This, in turn, suggests that
it is highly unlikely to observe significant values of \PhiR due to artefacts
induced by (uncorrelated observational) noise.

\section{Integrated information and distributed computation}
\label{sec:automata}

In the previous section we related integrated information to dynamical
complexity by linking \PhiR with criticality and metastability in coupled
oscillators. We now move on to Cellular Automata (CA), a well-known class of
systems widely used in the study of distributed
computation.\cite{mitchell1996computation,Lizier2010} Our aim here is to relate
IIT to distributed computation in two ways: at a global scale, \PhiR is higher
for complex, class \textsc{IV}\cite{Wolfram1984} automata; and at a local scale
\PhiR is higher for emergent coherent structures, like blinkers, gliders, and
collisions.

A CA is a multi-agent system in which every agent has a finite set of possible
states, and evolves in discrete time steps following a set of simple rules
based on its own and other agents' states. CA have been often used in studies
of self-organisation,\cite{Wolfram2002,Rosas2018} and some of them are capable
of universal computation.\cite{Wolfram1984} In a CA agents (or \emph{cells})
are arranged in a one-dimensional cyclic array (or \emph{tape}). The state of
each cell at a given time-step has a finite number of possible states, which is
determined via a boolean function (or \emph{rule}) which uses as arguments the
state of itself and its immediate neighbours at the previous time-step. The
same boolean function dictates the evolution of all agents in the system,
inducing a spatial translational symmetry. Each CA, irrespective of its number
of agents, is usually denoted by its rule.\cite{Wolfram2002}

For all the results presented below, we follow the simulation parameters used
by Lizier in his study of local information dynamics in CA:\cite{Lizier2010} we
initialise a tape of length \num{e4} with i.i.d. random variables, discard the
first 100 steps of simulation, and run 600 more steps that are used to estimate
the probability distributions used in all information-theoretic measures.

\begin{figure*}[ht!]
  \centering
  \includegraphics{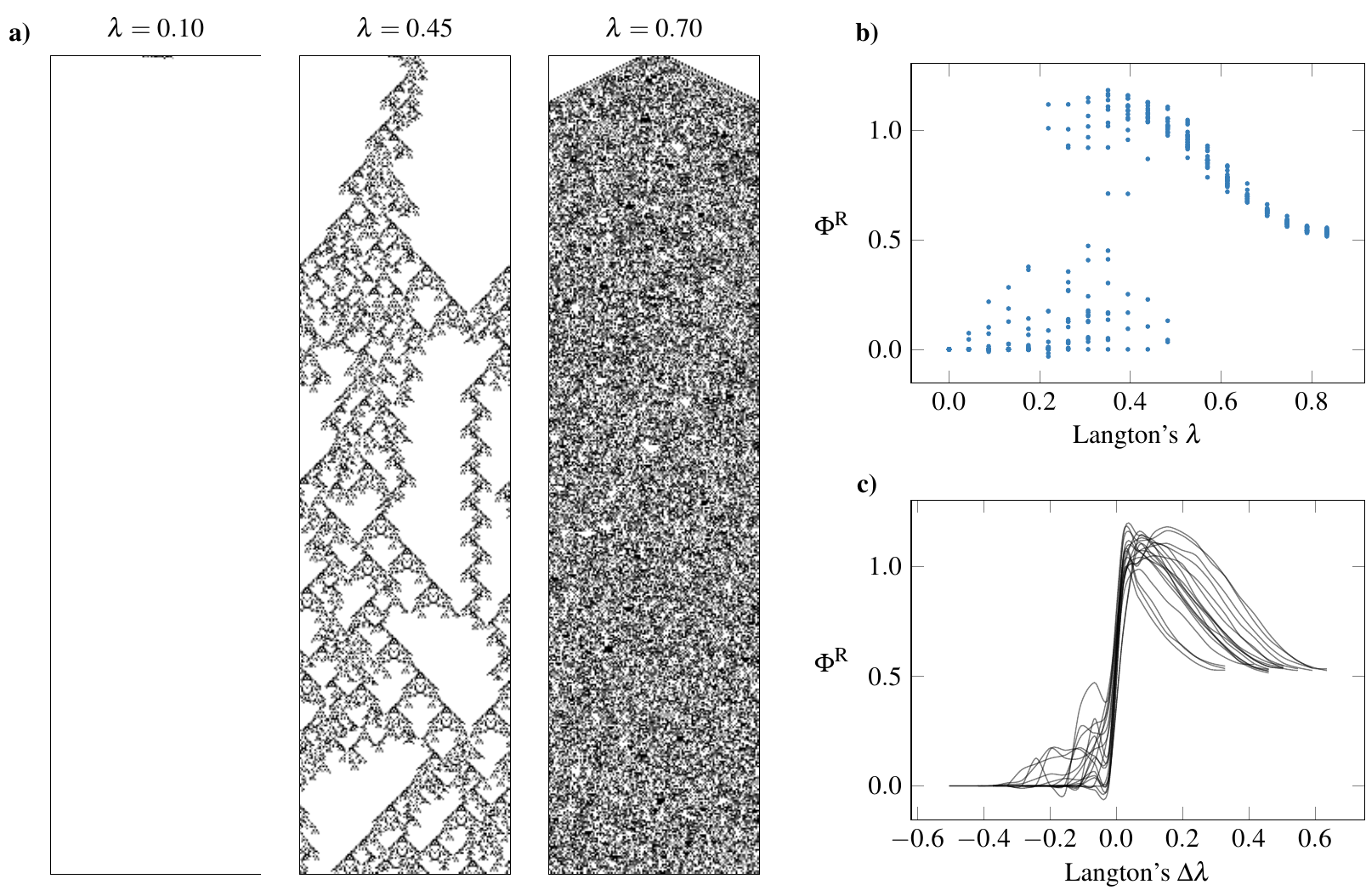}

\caption{\textbf{Integrated information peaks at the edge of chaos}.
  \textbf{(a)} Sample runs from a random 6-colour, range-2 cellular automaton
  with different $\lambda$ values, starting from a blank tape with 20
  randomised cells in the middle. The parameter $\lambda$ corresponds to the
  fraction of non-blank cells in an automaton's update rule. \textbf{(b)}
  Integrated information \PhiR peaks at an intermediate level of $\lambda$.
  \textbf{(c)} When plotted against $\Delta\lambda$, the distance from a
transition event, all runs align on a similar \PhiR profile.}

\label{fig:edgeofchaos}
\end{figure*}

\subsection{Integrated information and complexity classes}

Our first analysis focuses on Elementary Cellular Automata (ECA), a specific
subclass of CA. In ECA, each cell has two possible states (usually denoted as
white or black) ECA are traditionally denoted by their rule number, between 0
and 255, and grouped in four complexity classes:\cite{Wolfram1984} Class
\textsc{I} rules have attractors consisting of single absorbing states; Class
\textsc{II} rules evolve towards periodic orbits with relatively short cycle
lengths; and Class \textsc{III} and \textsc{IV} rules have attractors with
length of the order of the size of their phase space, with the latter being
characterised by the presence of highly structured patterns and persistent
structures.

As a first experiment, we calculate the average integrated information of each
ECA, separating each automaton by complexity class (Fig.~\ref{fig:phiwolfram}).
For this, we followed the classifications defined in Wolfram's original
article~\cite{Wolfram1984} as well as other clear-cut rules, and excluded
border cases which did not neatly fit into a single category.

Results show that \PhiR correlates strongly with complexity as discussed by
Wolfram: automata of higher classes have consistently higher \PhiR than
automata of lower classes, and the difference between classes \textsc{I,II} and
\textsc{III,IV} is stark.

It is worth noting the small difference between classes \textsc{III} and
\textsc{IV}. This is likely related to the blurriness of the line separating
both classes --- visually, it is hard to judge whether structures are
``coherent enough'' to support distributed computation and, formally, the
problem of determining whether a particular rule belongs to class III or IV is
considered undecidable.\cite{Culik1988,Martinez2013} Based on this, we may
tentatively suggest that the capacity to integrate information is a necessary,
but not sufficient, condition for universal computation.

\subsection{Integrated information at the edge of chaos}

In his seminal 1990 article, Langton\cite{Langton1990} took a step beyond
Wolfram's classification and argued that the complexity and universality
observed in ECA may reflect a broader phenomenon he called \emph{computation at
the edge of chaos}. In this view, computation is made possible by indefinitely
long transient states, a manifestation of \emph{critical
slowing-down}\cite{Scheffer2009} that form the particle-like structures seen in
class \textsc{IV} rules.

\begin{figure*}[ht!]
  \centering
  \includegraphics{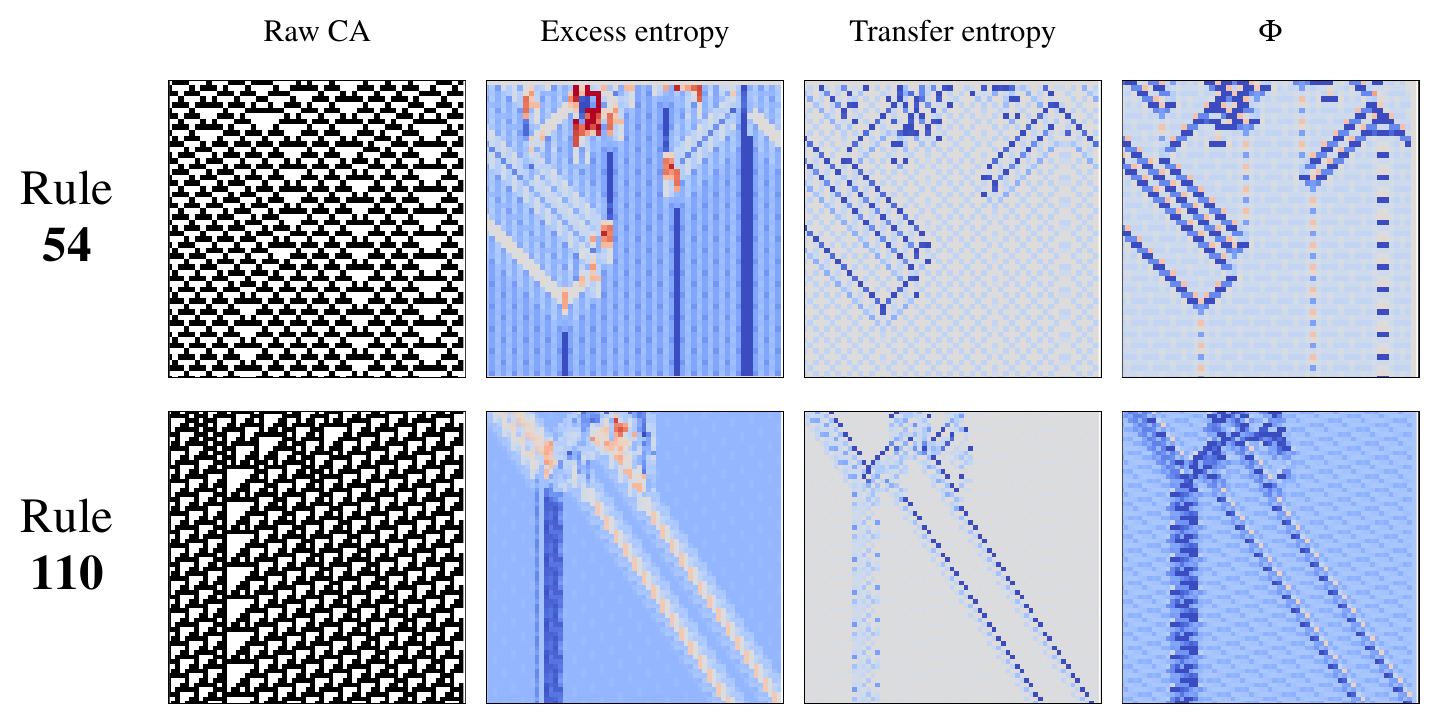}

  \caption{\textbf{Local integrated information detects coherent structures in
    cellular automata}.  Local information measures in cellular automata,
    applied to simulations of rule 54 (top) and 110 (bottom). Local excess
    entropy is high for static particles, local transfer entropy for moving
  particles, and local integrated information \phii for both. For all these
measures, blue and red indicate positive and negative values, respectively.}

  \label{fig:localinfoeca}
\end{figure*}

Langton's argument starts by defining a parameter $\lambda$, which represents
the fraction of neighbourhoods in a CA's rule table that map to a non-quiescent
state (i.e. a non-white colour). Then, by initialising one automaton with an
empty rule table and progressively filling it with non-quiescent states, one
can observe a transition point with exponentially long, particle-like
transients (Fig.~\subref{fig:edgeofchaos}{a}). Here we repeat Langton's
experiments using a 6-colour, range-2 CA, and compute its average \phii as its
rule table gets populated and $\lambda$ increases.

In agreement with Langton's argument, we found that integrated information has
largest values for intermediate values of $\lambda$, coinciding with the
automata's transition to a chaotic regime (Fig.~\subref{fig:edgeofchaos}{b}).
Interestingly, this shows that rules with high \PhiR are the ones at the
critical region --- where computation is possible.

Another unusual feature of Fig.~\subref{fig:edgeofchaos}{b} is that there is a
region where complex, high-\PhiR automata coexist with simpler ones. This
phenomenon was reported by Langton\cite{Langton1990} already: different
automata experience a ``transition to chaos'' at different values of $\lambda$.
This motivated a further analysis of measures of complexity as a function of
$\Delta\lambda$, the distance from the transition event for that particular
automaton. As expected, when aligned by their critical $\lambda$ value and
plotted against $\Delta\lambda$ (Fig.~\subref{fig:edgeofchaos}{c}), all curves
align onto a consistent, seemingly universal, picture of integrated information
across the $\lambda$ range.

For completeness, it is worth mentioning why at the right side of
Figs.~\subref{fig:edgeofchaos}{b} and~\subref{fig:edgeofchaos}{c} \PhiR does
not vanish for high $\lambda$ (as one could expect, given that the single-cell
autocorrelation does~\cite{Langton1990}). This is essentially due to the
determinism and locality of the automaton's rule: given a spatially extended
set of cells, it is always possible to predict the middle ones with perfect
certainty. At the same time, cutting the system with a bipartition will reduce
the spatial extent of this predictable region, so that the predictability of
the whole is greater than the predictability of the parts, and thus $\phii >
0$.

\subsection{Information is integrated by coherent structures}
\label{sec:structures}

In the experiments above we have shown that more complex automata integrate
more information. However, this is not enough to make a case for \PhiR as a
marker of distributed computation --- it may just be the case that
medium-$\lambda$ CA have higher \PhiR due to general properties of their rule
tables, or for some other reasons. In this section we address this possible
counter-argument by showing that the increase in \PhiR is due to the emerging
particles, and therefore can be directly associated with distributed
computation.

To show this, we run large simulations of ECA rules 54 and 110, and evaluate
several local information measures in small fragments of both
(Fig.~\ref{fig:localinfoeca}). Specifically, we compute Lizier's measures of
storage (excess entropy, $e_k$) and transfer (transfer entropy, $t_k$), as well
as local integrated information, using the binary time series of local
neighbourhoods of ECA cells as input (details in Appendix~\ref{app:local}).
Note that this pointwise \phii has not been $\Phi$ID-revised, as the
development of pointwise $\Phi$ID metrics is still an open
problem.\footnote{Indeed, although some pointwise redundancy functions
exist,\cite{Finn2018} the relatively much simpler redundancy in
Eq.~\eqref{eq:phiR} is not naturally translated into a pointwise setting, and
thus we restrict the pointwise analyses to standard integrated information
only.} A dedicated treatment of pointwise $\Phii$ID is part of our ongoing
work, and will be presented in a separate publication.

As expected, TE is high in gliders (moving particles), while excess entropy is
high in blinkers (static particles), confirming Lizier's results that these
structures perform information transfer and storage in CA.\cite{Lizier2010}
More interestingly for our purposes, is that \phii is high in \emph{all of
them} --- gliders, blinkers, and the collisions between them.

When studied at a local scale in space and time, we see that information
integration encompasses the three categories --- storage, transfer, and
modification --- and that \phii can detect all of them \emph{without having
been explicitly designed to do so}. This reinforces our claim that \phii (and,
in turn, its revised version \PhiR) is a generic marker of emergent dynamics,
and is connected with known measures of information processing. This
relationship can be understood in a more mathematically rigorous manner via
\phiid,\cite{Mediano2019a} which shows that \phii is decomposable in terms of
information transfer and synergy.

\section{Discussion}
\label{sec:discussion}

Many theoretical efforts to understand complexity have their roots in
information-theoretic or dynamical systems perspectives. On the one hand,
information-theoretic approaches focus on discrete systems displaying
coordinated activity capable of universal computation, e.g. via particles in
Wolfram's class \textsc{IV} automata.\cite{Wolfram1984} On the other hand,
dynamical approaches focus on continuous coupled dynamical systems near a phase
transition, displaying a stream of transient states of partial coherence that
balance robustness and adaptability.\cite{Kelso1995,rabinovich2008transient}
The results presented in this paper reveal how integrated information is
effective at characterising complexity in both information-processing and
dynamical systems, and hence can be regarded as a common signature of complex
behaviour across a wide range of systems. Overall, this paper shows how a
grounded, demystified understanding of IIT combined with the improved metrics
provided by $\Phi$ID can be used as a first step towards a much needed point of
contact between these diverging branches of complexity science, helping to
bridge the gap between information-processing and dynamical approaches.

\subsection{On the relationship between integrated information and
metastability, criticality, and distributed computation}

It is important to remark that the relationship between integrated information,
metastability, criticality, and distributed computation is not an identity, and
that their agreement is an important finding conveyed by $\Phi$.

Metastability, in the case of oscillator networks, is a community-local
quantity --- that is, it corresponds to an average over quantities
($\lambda_c$) that depend on the temporal diversity seen within each community,
independently of the rest. In stark contrast, $\Phi$ relies on the irreducible
interactions between communities. Interestingly, our finding reveals a close
relationship between the two, insofar as internal variability enables the
system to visit a larger repertoire of states in which system-wide interaction
can take place.

Criticality is, in general, enabled by a precise balance of two opposite forces
(typically characterised as order and chaos in physics) that enables peculiar
and fascinating phenomena such as scale-freeness, extreme sensitivity to
perturbations, and universality classes.\cite{beggs2012being,pruessner2012self}
In contrast, one of the core ideas in IIT is that integration and
differentiation are not opposite forces, but can actually coexist
together.\cite{Tononi1994,Rosas2019} Therefore, integrated information is not
maximised by optimal trade-offs, but by mechanisms that incentivise direct and
synergistic information transfer.\cite{Mediano2019a}

Finally, distributed computation is mainly based on intuitive but ultimately
informal notions developed after Wolfram's and Langton's work. Our results
establish a strong relationship between the capability of
information-processing complex systems to integrate information, and their
ability to perform highly non-trivial information processing via emergent
coherent structures, like blinkers, gliders, and collisions. The fact that
higher integrated information was carried by coherent emerging structures is
consistent with recent accounts of \emph{causally emergent
dynamics},\cite{rosas2020reconciling} which further supports the case of \phii
being an effective quantitative indicator of distributed emergent computation
with numerous potential future applications.

\subsection{Related work}

The conceptual link between complexity and integrated information is by no
means a novel idea: in fact, in the early days of IIT integrated information
and complexity were closely intertwined\cite{Tononi1998} --- as was its close
relation with the theory of complex networks.\cite{Sporns2000} Unfortunately,
as the theory turned more convoluted and less applicable, this link lost
relevance and IIT drifted away from standard information-theoretic methods.

Nonetheless, recent research has again brought complexity science to the fore
within the IIT community. In some cases, information-theoretic measures
originally conceived as measures of complexity have been re-purposed within
IIT.\cite{Ay2015} In others, new measures of complexity are inspired by
previous work on IIT.\cite{Langer2020} In contrast with its
consciousness-focussed sibling, advancements in this more pragmatic IIT (c.f.
Sec~\ref{sec:core}) have been enabled by a simplification and unification of
the underlying principles of the theory -- for example in terms of information
geometry\cite{Oizumi2016} or information decomposition.\cite{Mediano2019a}

Given its origins in theoretical neuroscience, it is no surprise that most of
IIT's relevance to complex systems has been mostly on models of neural
dynamics. In this context, the combination of integration and segregation
captured by \phii has been linked to other dynamical features of interest (like
metastability and neural avalanches) in a variety of settings, including e.g.
models of whole-brain activity\cite{Tagliazucchi2017} and spiking neural
networks.\cite{Mediano2017}

Finally, it is worth noting there has been recent work linking \phii with phase
transitions in thermodynamic systems.\cite{Aguilera2019} Together with recent
results linking information processing and stochastic
thermodynamics,\cite{parrondo2015thermodynamics} this exciting avenue of
research opens the door for a purely physical interpretation of \phii in
general physical systems.

\subsection{Concluding remarks}

This paper puts forward a pragmatic argument that metrics of integrated
information --- such as the ones provided by IIT and $\Phi$ID --- can allow us
to investigate unexplored commonalities between informational and dynamical
complexity, pointing out a promising avenue to reconcile their divide and
benefit subsidiary disciplines like cognitive and computational neuroscience.
Note that \phii is by no means the only quantity that peaks with the system's
complexity --- in cellular automata one could use the autocorrelation of a
single cell, and in coupled oscillators the variance of Kuramoto's order
parameter. However, the feature that makes \phii unique is that \emph{it is
applicable across the board}, and yields the desired results in different kinds
of systems without requiring idiosyncratic, ad-hoc measures. This unification,
analogous to the transversal role that Fisher information plays in phase
transitions over arbitrary order parameters,\cite{Prokopenko2011} posits
integrated information as a key quantity for the study of complex dynamics,
whose relevance has just started to be uncovered.

\section*{Acknowledgements}

P.M. and D.B. are funded by the Wellcome Trust (grant no. 210920/Z/18/Z). F.R.
is supported by the Ad Astra Chandaria foundation. A.B.B. is grateful to the
Dr. Mortimer and Theresa Sackler Foundation, which supports the Sackler Centre
for Consciousness Science.

\appendix

\section{Local integrated information}
\label{app:local}

A natural way to extend IIT 2.0 for complex systems analysis is to consider
local versions of $\Phi$, which can be built via the framework introduced by
Lizier.\cite{Lizier2007} Local (or pointwise) information measures are able to
identify coherent, emergent structures known as \emph{particles}, which have
been shown to be the basis of the distributed information processing that takes
place in systems such as cellular
automata.\cite{Lizier2010,Lizier2012,Lizier2010c}

One of the most basic pointwise information metrics is the local mutual
information, which is defined as
\begin{align}
  i(x; y) := \log \frac{p(x, y)}{p(x) p(y)} ~ ,
  \label{eq:localmi}
\end{align}
\noindent so that $\mathbb{E}[i(X; Y)] = I(X;Y)$ is the usual mutual
information. By evaluating $i$ on every $x,y$ pair, one can determine which
particular combinations of symbols play a predominant role for the observed
interdependency between $X$ and $Y$. (More specifically, the local
mutual information captures specific deviations between the joint distribution
and the product of the marginals.) Building on these ideas, Lizier proposed a
taxonomy of distributed information processing as composed of \emph{storage},
\emph{transfer} and \emph{modification}.\cite{Lizier2010} For this, consider a
bivariate stochastic process $(X_t,Y_t)$ with $t\in\mathbb{Z}$, and introduce
the shorthand notation $X^{(k)}_t = (X_{t-k}, \dots, X_t)$ and $X^{(k^+)}_t =
(X_t,\dots , X_{t+k-1})$ for the corresponding past and future embedding vectors
of length $k$. In this context, storage within the subprocess $X_t$ is
identified with its \emph{excess entropy} $\mathrm{E}_k =I(X^{(k)}_t;
X^{(k^+)}_{t+1})$,\cite{Crutchfield2003} and transfer from $Y_t$ to $X_{t+1}$
with the \emph{transfer entropy} $\mathrm{TE}_k = I(X_{t+1} ; X^{(k)}_t |
Y^{(k)}_t)$.\cite{Schreiber2000} Interestingly, both quantities have
corresponding local versions:
\begin{align}
  e_k(x_t)      &:= \log \frac{p(x^{(k)}_t, x^{(k^+)}_{t+1})}{p(x^{(k)}_t) p(x^{(k^+)}_{t+1})} ~ , \\
  t_k(y_t \rightarrow x_t) &:= \log \frac{p(x_{t+1} | x^{(k)}_t, y^{(k)}_t)}{p(x_{t+1} | x^{(k)}_t)} ~ ,
  \label{eq:localinfo}
\end{align}
\noindent such that, as expected, $\mathbb{E}[e_k] = \mathrm{E}_k$ and
$\mathbb{E}[t_k] = \mathrm{TE}_k$.
Note that to measure transfer in either
direction for the results in Fig.~\ref{fig:localinfoeca}, we compute the 
local TE from a cell to its left and right neighbours and take the maximum 
of the two.

These ideas can be used to extend the standard formulation of integrated
information measures in two ways. First, by using embedding vectors, the IIT
metrics are applicable to non-Markovian systems. Second, by formulating
pointwise measures one can capture spatio-temporal variations in
\Phii. Mathematically, we reformulate
Eq.~\eqref{eq:phi} introducing these modifications as
\begin{align}
  \varphi_k[X; \tau, \mathcal{B}] = I(X^{(k)}_{t-\tau}; X_t) - \displaystyle\sum_{j=1}^2 I(M^{j,(k)}_{t-\tau}; M^{j}_t) ~ ,
  \label{eq:phiembedding}
\end{align}
\noindent and apply the same partition scheme described in
Sec.~\ref{sec:math_behind_phi} to obtain an `embedded' integrated information,
$\phii_k$. Then, the equation above can be readily made into a local measure by
replacing mutual information with its local counterpart,
\begin{align}
  \phi_k[x_t; \tau, \mathcal{B}] = i(x^{(k)}_{t-\tau}; x_t) - \displaystyle\sum_{j=1}^2 i(m^{j,(k)}_{t-\tau}; m^{j}_t) ~ ,
  \label{eq:localphi}
\end{align}
\noindent such that, as expected, $\varphi_k[X; \tau, \mathcal{B}] =
\mathbb{E}\left[ \phi_k[x_t; \tau, \mathcal{B}] \right]$.

\bibliography{main.bib}

\begin{thebibliography}{70}%
\makeatletter
\providecommand \@ifxundefined [1]{%
 \@ifx{#1\undefined}
}%
\providecommand \@ifnum [1]{%
 \ifnum #1\expandafter \@firstoftwo
 \else \expandafter \@secondoftwo
 \fi
}%
\providecommand \@ifx [1]{%
 \ifx #1\expandafter \@firstoftwo
 \else \expandafter \@secondoftwo
 \fi
}%
\providecommand \natexlab [1]{#1}%
\providecommand \enquote  [1]{``#1''}%
\providecommand \bibnamefont  [1]{#1}%
\providecommand \bibfnamefont [1]{#1}%
\providecommand \citenamefont [1]{#1}%
\providecommand \href@noop [0]{\@secondoftwo}%
\providecommand \href [0]{\begingroup \@sanitize@url \@href}%
\providecommand \@href[1]{\@@startlink{#1}\@@href}%
\providecommand \@@href[1]{\endgroup#1\@@endlink}%
\providecommand \@sanitize@url [0]{\catcode `\\12\catcode `\$12\catcode
  `\&12\catcode `\#12\catcode `\^12\catcode `\_12\catcode `\%12\relax}%
\providecommand \@@startlink[1]{}%
\providecommand \@@endlink[0]{}%
\providecommand \url  [0]{\begingroup\@sanitize@url \@url }%
\providecommand \@url [1]{\endgroup\@href {#1}{\urlprefix }}%
\providecommand \urlprefix  [0]{URL }%
\providecommand \Eprint [0]{\href }%
\providecommand \doibase [0]{http://dx.doi.org/}%
\providecommand \selectlanguage [0]{\@gobble}%
\providecommand \bibinfo  [0]{\@secondoftwo}%
\providecommand \bibfield  [0]{\@secondoftwo}%
\providecommand \translation [1]{[#1]}%
\providecommand \BibitemOpen [0]{}%
\providecommand \bibitemStop [0]{}%
\providecommand \bibitemNoStop [0]{.\EOS\space}%
\providecommand \EOS [0]{\spacefactor3000\relax}%
\providecommand \BibitemShut  [1]{\csname bibitem#1\endcsname}%
\let\auto@bib@innerbib\@empty
\bibitem [{\citenamefont {Wolfram}(2002)}]{Wolfram2002}%
  \BibitemOpen
  \bibfield  {author} {\bibinfo {author} {\bibfnamefont {S.}~\bibnamefont
  {Wolfram}},\ }\href@noop {} {\emph {\bibinfo {title} {{A New Kind of
  Science}}}}\ (\bibinfo  {publisher} {Wolfram Media},\ \bibinfo {year}
  {2002})\ p.\ \bibinfo {pages} {1197}\BibitemShut {NoStop}%
\bibitem [{\citenamefont {Pikovsky}, \citenamefont {Rosenblum},\ and\
  \citenamefont {Kurths}(2001)}]{Pikovsky2001}%
  \BibitemOpen
  \bibfield  {author} {\bibinfo {author} {\bibfnamefont {A.}~\bibnamefont
  {Pikovsky}}, \bibinfo {author} {\bibfnamefont {M.}~\bibnamefont {Rosenblum}},
  \ and\ \bibinfo {author} {\bibfnamefont {J.}~\bibnamefont {Kurths}},\
  }\href@noop {} {\emph {\bibinfo {title} {{Synchronization: A Universal
  Concept in Nonlinear Sciences}}}}\ (\bibinfo  {publisher} {Cambridge
  University Press},\ \bibinfo {year} {2001})\ p.\ \bibinfo {pages}
  {432}\BibitemShut {NoStop}%
\bibitem [{\citenamefont {Fodor}(1975)}]{fodor1975language}%
  \BibitemOpen
  \bibfield  {author} {\bibinfo {author} {\bibfnamefont {J.~A.}\ \bibnamefont
  {Fodor}},\ }\href@noop {} {\emph {\bibinfo {title} {The Language of
  Thought}}},\ Vol.~\bibinfo {volume} {5}\ (\bibinfo  {publisher} {Harvard
  University Press},\ \bibinfo {year} {1975})\BibitemShut {NoStop}%
\bibitem [{\citenamefont {Pylyshyn}\ and\ \citenamefont
  {Turnbiull}(1986)}]{pylyshyn1986computation}%
  \BibitemOpen
  \bibfield  {author} {\bibinfo {author} {\bibfnamefont {Z.}~\bibnamefont
  {Pylyshyn}}\ and\ \bibinfo {author} {\bibfnamefont {W.}~\bibnamefont
  {Turnbiull}},\ }\bibfield  {title} {\enquote {\bibinfo {title} {Computation
  and cognition: Toward a foundation for cognitive science},}\ }\href@noop {}
  {\bibfield  {journal} {\bibinfo  {journal} {Canadian Psychology}\ }\textbf
  {\bibinfo {volume} {27}},\ \bibinfo {pages} {85--87} (\bibinfo {year}
  {1986})}\BibitemShut {NoStop}%
\bibitem [{\citenamefont {Rescorla}(2017)}]{rescorla2017ockham}%
  \BibitemOpen
  \bibfield  {author} {\bibinfo {author} {\bibfnamefont {M.}~\bibnamefont
  {Rescorla}},\ }\bibfield  {title} {\enquote {\bibinfo {title} {From {Ockham}
  to {Turing} -- and back again},}\ }in\ \href {\doibase
  10.1007/978-3-319-53280-6_12} {\emph {\bibinfo {booktitle} {Philosophical
  Explorations of the Legacy of Alan Turing}}}\ (\bibinfo  {publisher}
  {Springer},\ \bibinfo {year} {2017})\ pp.\ \bibinfo {pages}
  {279--304}\BibitemShut {NoStop}%
\bibitem [{\citenamefont {Milkowski}(2013)}]{milkowski2013explaining}%
  \BibitemOpen
  \bibfield  {author} {\bibinfo {author} {\bibfnamefont {M.}~\bibnamefont
  {Milkowski}},\ }\href@noop {} {\emph {\bibinfo {title} {Explaining the
  Computational Mind}}}\ (\bibinfo  {publisher} {MIT Press},\ \bibinfo {year}
  {2013})\BibitemShut {NoStop}%
\bibitem [{\citenamefont {Van~Gelder}(1995)}]{van1995might}%
  \BibitemOpen
  \bibfield  {author} {\bibinfo {author} {\bibfnamefont {T.}~\bibnamefont
  {Van~Gelder}},\ }\bibfield  {title} {\enquote {\bibinfo {title} {What might
  cognition be, if not computation?}}\ }\href {\doibase 10.2307/2941061}
  {\bibfield  {journal} {\bibinfo  {journal} {The Journal of Philosophy}\
  }\textbf {\bibinfo {volume} {92}},\ \bibinfo {pages} {345--381} (\bibinfo
  {year} {1995})}\BibitemShut {NoStop}%
\bibitem [{\citenamefont {Van~Gelder}(1998)}]{van1998dynamical}%
  \BibitemOpen
  \bibfield  {author} {\bibinfo {author} {\bibfnamefont {T.}~\bibnamefont
  {Van~Gelder}},\ }\bibfield  {title} {\enquote {\bibinfo {title} {The
  dynamical hypothesis in cognitive science},}\ }\href {\doibase
  10.1017/S0140525X98001733} {\bibfield  {journal} {\bibinfo  {journal}
  {Behavioral and Brain Sciences}\ }\textbf {\bibinfo {volume} {21}},\ \bibinfo
  {pages} {615--628} (\bibinfo {year} {1998})}\BibitemShut {NoStop}%
\bibitem [{\citenamefont {Smith}(2005)}]{smith2005cognition}%
  \BibitemOpen
  \bibfield  {author} {\bibinfo {author} {\bibfnamefont {L.~B.}\ \bibnamefont
  {Smith}},\ }\bibfield  {title} {\enquote {\bibinfo {title} {Cognition as a
  dynamic system: Principles from embodiment},}\ }\href {\doibase
  10.1016/j.dr.2005.11.001} {\bibfield  {journal} {\bibinfo  {journal}
  {Developmental Review}\ }\textbf {\bibinfo {volume} {25}},\ \bibinfo {pages}
  {278--298} (\bibinfo {year} {2005})}\BibitemShut {NoStop}%
\bibitem [{\citenamefont {Sch{\"o}ner}(2008)}]{schoner2008dynamical}%
  \BibitemOpen
  \bibfield  {author} {\bibinfo {author} {\bibfnamefont {G.}~\bibnamefont
  {Sch{\"o}ner}},\ }\bibfield  {title} {\enquote {\bibinfo {title} {Dynamical
  systems approaches to cognition},}\ }\href {\doibase
  10.1017/CBO9780511816772.007} {\bibfield  {journal} {\bibinfo  {journal}
  {Cambridge Handbook of Computational Cognitive Modeling}\ ,\ \bibinfo {pages}
  {101--126}} (\bibinfo {year} {2008})}\BibitemShut {NoStop}%
\bibitem [{\citenamefont {Minsky}(1991)}]{minsky1991logical}%
  \BibitemOpen
  \bibfield  {author} {\bibinfo {author} {\bibfnamefont {M.~L.}\ \bibnamefont
  {Minsky}},\ }\bibfield  {title} {\enquote {\bibinfo {title} {Logical versus
  analogical or symbolic versus connectionist or neat versus scruffy},}\
  }\href@noop {} {\bibfield  {journal} {\bibinfo  {journal} {AI magazine}\
  }\textbf {\bibinfo {volume} {12}},\ \bibinfo {pages} {34--34} (\bibinfo
  {year} {1991})}\BibitemShut {NoStop}%
\bibitem [{\citenamefont {Garnelo}\ and\ \citenamefont
  {Shanahan}(2019)}]{Garnelo2019}%
  \BibitemOpen
  \bibfield  {author} {\bibinfo {author} {\bibfnamefont {M.}~\bibnamefont
  {Garnelo}}\ and\ \bibinfo {author} {\bibfnamefont {M.}~\bibnamefont
  {Shanahan}},\ }\bibfield  {title} {\enquote {\bibinfo {title} {Reconciling
  deep learning with symbolic artificial intelligence: Representing objects and
  relations},}\ }\href {\doibase https://doi.org/10.1016/j.cobeha.2018.12.010}
  {\bibfield  {journal} {\bibinfo  {journal} {Current Opinion in Behavioral
  Sciences}\ }\textbf {\bibinfo {volume} {29}},\ \bibinfo {pages} {17--23}
  (\bibinfo {year} {2019})}\BibitemShut {NoStop}%
\bibitem [{\citenamefont {Beer}\ and\ \citenamefont
  {Williams}(2015)}]{beer2015information}%
  \BibitemOpen
  \bibfield  {author} {\bibinfo {author} {\bibfnamefont {R.~D.}\ \bibnamefont
  {Beer}}\ and\ \bibinfo {author} {\bibfnamefont {P.~L.}\ \bibnamefont
  {Williams}},\ }\bibfield  {title} {\enquote {\bibinfo {title} {Information
  processing and dynamics in minimally cognitive agents},}\ }\href {\doibase
  10.1111/cogs.12142} {\bibfield  {journal} {\bibinfo  {journal} {Cognitive
  Science}\ }\textbf {\bibinfo {volume} {39}},\ \bibinfo {pages} {1--38}
  (\bibinfo {year} {2015})}\BibitemShut {NoStop}%
\bibitem [{\citenamefont {Wolfram}(1984)}]{Wolfram1984}%
  \BibitemOpen
  \bibfield  {author} {\bibinfo {author} {\bibfnamefont {S.}~\bibnamefont
  {Wolfram}},\ }\bibfield  {title} {\enquote {\bibinfo {title} {Universality
  and complexity in cellular automata},}\ }\href {\doibase
  10.1016/0167-2789(84)90245-8} {\bibfield  {journal} {\bibinfo  {journal}
  {Physica D}\ }\textbf {\bibinfo {volume} {10}},\ \bibinfo {pages} {1--35}
  (\bibinfo {year} {1984})}\BibitemShut {NoStop}%
\bibitem [{\citenamefont {Langton}(1990)}]{Langton1990}%
  \BibitemOpen
  \bibfield  {author} {\bibinfo {author} {\bibfnamefont {C.~G.}\ \bibnamefont
  {Langton}},\ }\bibfield  {title} {\enquote {\bibinfo {title} {Computation at
  the edge of chaos: {Phase} transitions and emergent computation},}\ }\href
  {\doibase 10.1016/0167-2789(90)90064-V} {\bibfield  {journal} {\bibinfo
  {journal} {Physica D: Nonlinear Phenomena}\ }\textbf {\bibinfo {volume}
  {42}},\ \bibinfo {pages} {12--37} (\bibinfo {year} {1990})}\BibitemShut
  {NoStop}%
\bibitem [{\citenamefont {Tononi}\ and\ \citenamefont
  {Sporns}(2003)}]{Tononi2003}%
  \BibitemOpen
  \bibfield  {author} {\bibinfo {author} {\bibfnamefont {G.}~\bibnamefont
  {Tononi}}\ and\ \bibinfo {author} {\bibfnamefont {O.}~\bibnamefont
  {Sporns}},\ }\bibfield  {title} {\enquote {\bibinfo {title} {Measuring
  information integration},}\ }\href {\doibase 10.1186/1471-2202-4-31}
  {\bibfield  {journal} {\bibinfo  {journal} {BMC Neuroscience}\ }\textbf
  {\bibinfo {volume} {4}},\ \bibinfo {pages} {31} (\bibinfo {year}
  {2003})}\BibitemShut {NoStop}%
\bibitem [{\citenamefont {Balduzzi}\ and\ \citenamefont
  {Tononi}(2008)}]{Balduzzi2008}%
  \BibitemOpen
  \bibfield  {author} {\bibinfo {author} {\bibfnamefont {D.}~\bibnamefont
  {Balduzzi}}\ and\ \bibinfo {author} {\bibfnamefont {G.}~\bibnamefont
  {Tononi}},\ }\bibfield  {title} {\enquote {\bibinfo {title} {Integrated
  information in discrete dynamical dystems: {Motivation} and theoretical
  framework},}\ }\href {\doibase 10.1371/journal.pcbi.1000091} {\bibfield
  {journal} {\bibinfo  {journal} {PLoS Computational Biology}\ }\textbf
  {\bibinfo {volume} {4}},\ \bibinfo {pages} {e1000091} (\bibinfo {year}
  {2008})}\BibitemShut {NoStop}%
\bibitem [{\citenamefont {Oizumi}, \citenamefont {Albantakis},\ and\
  \citenamefont {Tononi}(2014)}]{Oizumi2014}%
  \BibitemOpen
  \bibfield  {author} {\bibinfo {author} {\bibfnamefont {M.}~\bibnamefont
  {Oizumi}}, \bibinfo {author} {\bibfnamefont {L.}~\bibnamefont {Albantakis}},
  \ and\ \bibinfo {author} {\bibfnamefont {G.}~\bibnamefont {Tononi}},\
  }\bibfield  {title} {\enquote {\bibinfo {title} {From the phenomenology to
  the mechanisms of consciousness: {Integrated Information Theory} 3.0},}\
  }\href {\doibase 10.1371/journal.pcbi.1003588} {\bibfield  {journal}
  {\bibinfo  {journal} {PLoS Computational Biology}\ }\textbf {\bibinfo
  {volume} {10}},\ \bibinfo {pages} {e1003588} (\bibinfo {year}
  {2014})}\BibitemShut {NoStop}%
\bibitem [{\citenamefont {Mediano}\ \emph {et~al.}(2019)\citenamefont
  {Mediano}, \citenamefont {Rosas}, \citenamefont {Carhart-Harris},
  \citenamefont {Seth},\ and\ \citenamefont {Barrett}}]{Mediano2019a}%
  \BibitemOpen
  \bibfield  {author} {\bibinfo {author} {\bibfnamefont {P.}~\bibnamefont
  {Mediano}}, \bibinfo {author} {\bibfnamefont {F.}~\bibnamefont {Rosas}},
  \bibinfo {author} {\bibfnamefont {R.~L.}\ \bibnamefont {Carhart-Harris}},
  \bibinfo {author} {\bibfnamefont {A.~K.}\ \bibnamefont {Seth}}, \ and\
  \bibinfo {author} {\bibfnamefont {A.~B.}\ \bibnamefont {Barrett}},\
  }\bibfield  {title} {\enquote {\bibinfo {title} {Beyond integrated
  information: {A} taxonomy of information dynamics phenomena},}\ }\href@noop
  {} {\  (\bibinfo {year} {2019})},\ \Eprint {http://arxiv.org/abs/1909.02297}
  {arXiv:1909.02297} \BibitemShut {NoStop}%
\bibitem [{\citenamefont {Tononi}\ and\ \citenamefont
  {Koch}(2015)}]{Tononi2015}%
  \BibitemOpen
  \bibfield  {author} {\bibinfo {author} {\bibfnamefont {G.}~\bibnamefont
  {Tononi}}\ and\ \bibinfo {author} {\bibfnamefont {C.}~\bibnamefont {Koch}},\
  }\bibfield  {title} {\enquote {\bibinfo {title} {Consciousness: {Here,} there
  and everywhere?}}\ }\href {\doibase 10.1098/rstb.2014.0167} {\bibfield
  {journal} {\bibinfo  {journal} {Philosophical Transactions of the Royal
  Society B}\ }\textbf {\bibinfo {volume} {370}},\ \bibinfo {pages} {20140167}
  (\bibinfo {year} {2015})}\BibitemShut {NoStop}%
\bibitem [{\citenamefont {Tononi}\ \emph {et~al.}(2016)\citenamefont {Tononi},
  \citenamefont {Boly}, \citenamefont {Massimini},\ and\ \citenamefont
  {Koch}}]{Tononi2016}%
  \BibitemOpen
  \bibfield  {author} {\bibinfo {author} {\bibfnamefont {G.}~\bibnamefont
  {Tononi}}, \bibinfo {author} {\bibfnamefont {M.}~\bibnamefont {Boly}},
  \bibinfo {author} {\bibfnamefont {M.}~\bibnamefont {Massimini}}, \ and\
  \bibinfo {author} {\bibfnamefont {C.}~\bibnamefont {Koch}},\ }\bibfield
  {title} {\enquote {\bibinfo {title} {Integrated information theory: {From}
  consciousness to its physical substrate},}\ }\href {\doibase
  10.1038/nrn.2016.44} {\bibfield  {journal} {\bibinfo  {journal} {Nature
  Reviews Neuroscience}\ }\textbf {\bibinfo {volume} {17}},\ \bibinfo {pages}
  {450} (\bibinfo {year} {2016})}\BibitemShut {NoStop}%
\bibitem [{\citenamefont {Barrett}\ and\ \citenamefont
  {Mediano}(2019)}]{Barrett2019}%
  \BibitemOpen
  \bibfield  {author} {\bibinfo {author} {\bibfnamefont {A.~B.}\ \bibnamefont
  {Barrett}}\ and\ \bibinfo {author} {\bibfnamefont {P.}~\bibnamefont
  {Mediano}},\ }\bibfield  {title} {\enquote {\bibinfo {title} {The {Phi}
  measure of integrated information is not well-defined for general physical
  systems},}\ }\href
  {https://www.ingentaconnect.com/content/imp/jcs/2019/00000026/f0020001/art00002}
  {\bibfield  {journal} {\bibinfo  {journal} {Journal of Consciousness
  Studies}\ }\textbf {\bibinfo {volume} {26}},\ \bibinfo {pages} {11--20}
  (\bibinfo {year} {2019})}\BibitemShut {NoStop}%
\bibitem [{\citenamefont {Cerullo}(2015)}]{Cerullo2015}%
  \BibitemOpen
  \bibfield  {author} {\bibinfo {author} {\bibfnamefont {M.~A.}\ \bibnamefont
  {Cerullo}},\ }\bibfield  {title} {\enquote {\bibinfo {title} {The problem
  with {Phi}: {A} critique of integrated information theory},}\ }\href
  {\doibase 10.1371/journal.pcbi.1004286} {\bibfield  {journal} {\bibinfo
  {journal} {PLoS Computational Biology}\ }\textbf {\bibinfo {volume} {11}}
  (\bibinfo {year} {2015}),\ 10.1371/journal.pcbi.1004286}\BibitemShut
  {NoStop}%
\bibitem [{\citenamefont {Mindt}(2017)}]{Mindt2017}%
  \BibitemOpen
  \bibfield  {author} {\bibinfo {author} {\bibfnamefont {G.}~\bibnamefont
  {Mindt}},\ }\bibfield  {title} {\enquote {\bibinfo {title} {The problem with
  the 'information' in integrated information theory},}\ }\href
  {https://www.ingentaconnect.com/contentone/imp/jcs/2017/00000024/F0020007/art00007}
  {\bibfield  {journal} {\bibinfo  {journal} {Journal of Consciousness
  Studies}\ }\textbf {\bibinfo {volume} {24}},\ \bibinfo {pages} {130--154}
  (\bibinfo {year} {2017})}\BibitemShut {NoStop}%
\bibitem [{\citenamefont {Anderson}(1972)}]{anderson1972more}%
  \BibitemOpen
  \bibfield  {author} {\bibinfo {author} {\bibfnamefont {P.~W.}\ \bibnamefont
  {Anderson}},\ }\bibfield  {title} {\enquote {\bibinfo {title} {More is
  different},}\ }\href {\doibase 10.1126/science.177.4047.393} {\bibfield
  {journal} {\bibinfo  {journal} {Science}\ }\textbf {\bibinfo {volume}
  {177}},\ \bibinfo {pages} {393--396} (\bibinfo {year} {1972})}\BibitemShut
  {NoStop}%
\bibitem [{\citenamefont {Waldrop}(1993)}]{Waldrop1993}%
  \BibitemOpen
  \bibfield  {author} {\bibinfo {author} {\bibfnamefont {M.~M.}\ \bibnamefont
  {Waldrop}},\ }\href@noop {} {\emph {\bibinfo {title} {Complexity: The
  Emerging Science at the Edge of Order and Chaos}}}\ (\bibinfo  {publisher}
  {Simon and Schuster},\ \bibinfo {year} {1993})\BibitemShut {NoStop}%
\bibitem [{\citenamefont {Barrett}\ and\ \citenamefont
  {Seth}(2011)}]{Barrett2011}%
  \BibitemOpen
  \bibfield  {author} {\bibinfo {author} {\bibfnamefont {A.~B.}\ \bibnamefont
  {Barrett}}\ and\ \bibinfo {author} {\bibfnamefont {A.~K.}\ \bibnamefont
  {Seth}},\ }\bibfield  {title} {\enquote {\bibinfo {title} {Practical measures
  of integrated information for time-series data},}\ }\href {\doibase
  10.1371/journal.pcbi.1001052} {\bibfield  {journal} {\bibinfo  {journal}
  {PLoS Computational Biology}\ }\textbf {\bibinfo {volume} {7}},\ \bibinfo
  {pages} {e1001052} (\bibinfo {year} {2011})}\BibitemShut {NoStop}%
\bibitem [{\citenamefont {Mediano}, \citenamefont {Seth},\ and\ \citenamefont
  {Barrett}(2019)}]{Mediano2019}%
  \BibitemOpen
  \bibfield  {author} {\bibinfo {author} {\bibfnamefont {P.}~\bibnamefont
  {Mediano}}, \bibinfo {author} {\bibfnamefont {A.~K.}\ \bibnamefont {Seth}}, \
  and\ \bibinfo {author} {\bibfnamefont {A.~B.}\ \bibnamefont {Barrett}},\
  }\bibfield  {title} {\enquote {\bibinfo {title} {Measuring integrated
  information: Comparison of candidate measures in theory and simulation},}\
  }\href {\doibase 10.3390/e21010017} {\bibfield  {journal} {\bibinfo
  {journal} {Entropy}\ }\textbf {\bibinfo {volume} {21}},\ \bibinfo {pages}
  {17} (\bibinfo {year} {2019})}\BibitemShut {NoStop}%
\bibitem [{\citenamefont {Barrett}(2015)}]{Barrett2015}%
  \BibitemOpen
  \bibfield  {author} {\bibinfo {author} {\bibfnamefont {A.~B.}\ \bibnamefont
  {Barrett}},\ }\bibfield  {title} {\enquote {\bibinfo {title} {Exploration of
  synergistic and redundant information sharing in static and dynamical
  {Gaussian} systems},}\ }\href {\doibase 10.1103/PhysRevE.91.052802}
  {\bibfield  {journal} {\bibinfo  {journal} {Physical Review E}\ }\textbf
  {\bibinfo {volume} {91}},\ \bibinfo {pages} {052802} (\bibinfo {year}
  {2015})}\BibitemShut {NoStop}%
\bibitem [{\citenamefont {Luppi}\ \emph {et~al.}(2020)\citenamefont {Luppi},
  \citenamefont {Mediano}, \citenamefont {Rosas}, \citenamefont {Allanson},
  \citenamefont {Pickard}, \citenamefont {Carhart-Harris}, \citenamefont
  {Williams}, \citenamefont {Craig}, \citenamefont {Finoia}, \citenamefont
  {Owen} \emph {et~al.}}]{luppi2020synergistic}%
  \BibitemOpen
  \bibfield  {author} {\bibinfo {author} {\bibfnamefont {A.~I.}\ \bibnamefont
  {Luppi}}, \bibinfo {author} {\bibfnamefont {P.~A.}\ \bibnamefont {Mediano}},
  \bibinfo {author} {\bibfnamefont {F.~E.}\ \bibnamefont {Rosas}}, \bibinfo
  {author} {\bibfnamefont {J.}~\bibnamefont {Allanson}}, \bibinfo {author}
  {\bibfnamefont {J.~D.}\ \bibnamefont {Pickard}}, \bibinfo {author}
  {\bibfnamefont {R.~L.}\ \bibnamefont {Carhart-Harris}}, \bibinfo {author}
  {\bibfnamefont {G.~B.}\ \bibnamefont {Williams}}, \bibinfo {author}
  {\bibfnamefont {M.~M.}\ \bibnamefont {Craig}}, \bibinfo {author}
  {\bibfnamefont {P.}~\bibnamefont {Finoia}}, \bibinfo {author} {\bibfnamefont
  {A.~M.}\ \bibnamefont {Owen}},  \emph {et~al.},\ }\bibfield  {title}
  {\enquote {\bibinfo {title} {A synergistic workspace for human consciousness
  revealed by integrated information decomposition},}\ }\href@noop {}
  {\bibfield  {journal} {\bibinfo  {journal} {bioRxiv}\ } (\bibinfo {year}
  {2020})}\BibitemShut {NoStop}%
\bibitem [{\citenamefont {Lizier}(2010)}]{Lizier2010}%
  \BibitemOpen
  \bibfield  {author} {\bibinfo {author} {\bibfnamefont {J.~T.}\ \bibnamefont
  {Lizier}},\ }\href {\doibase 10.1007/978-3-642-32952-4} {\emph {\bibinfo
  {title} {{The Local Information Dynamics of Distributed Computation in
  Complex Systems}}}},\ Springer Theses\ (\bibinfo  {publisher} {Springer},\
  \bibinfo {address} {Berlin, Heidelberg},\ \bibinfo {year} {2010})\BibitemShut
  {NoStop}%
\bibitem [{\citenamefont {Lizier}, \citenamefont {Prokopenko},\ and\
  \citenamefont {Zomaya}(2007)}]{Lizier2007}%
  \BibitemOpen
  \bibfield  {author} {\bibinfo {author} {\bibfnamefont {J.~T.}\ \bibnamefont
  {Lizier}}, \bibinfo {author} {\bibfnamefont {M.}~\bibnamefont {Prokopenko}},
  \ and\ \bibinfo {author} {\bibfnamefont {A.~Y.}\ \bibnamefont {Zomaya}},\
  }\bibfield  {title} {\enquote {\bibinfo {title} {Information transfer by
  particles in cellular automata},}\ }in\ \href {\doibase
  10.1007/978-3-540-76931-6_5} {\emph {\bibinfo {booktitle} {Australian
  Conference on Artificial Life}}}\ (\bibinfo {organization} {Springer},\
  \bibinfo {year} {2007})\ pp.\ \bibinfo {pages} {49--60}\BibitemShut {NoStop}%
\bibitem [{\citenamefont {Lizier}, \citenamefont {Prokopenko},\ and\
  \citenamefont {Zomaya}(2012)}]{Lizier2012}%
  \BibitemOpen
  \bibfield  {author} {\bibinfo {author} {\bibfnamefont {J.~T.}\ \bibnamefont
  {Lizier}}, \bibinfo {author} {\bibfnamefont {M.}~\bibnamefont {Prokopenko}},
  \ and\ \bibinfo {author} {\bibfnamefont {A.~Y.}\ \bibnamefont {Zomaya}},\
  }\bibfield  {title} {\enquote {\bibinfo {title} {Local measures of
  information storage in complex distributed computation},}\ }\href {\doibase
  10.1016/j.ins.2012.04.016} {\bibfield  {journal} {\bibinfo  {journal}
  {Information Sciences}\ }\textbf {\bibinfo {volume} {208}},\ \bibinfo {pages}
  {39--54} (\bibinfo {year} {2012})}\BibitemShut {NoStop}%
\bibitem [{\citenamefont {Lizier}, \citenamefont {Prokopenko},\ and\
  \citenamefont {Zomaya}(2010)}]{Lizier2010c}%
  \BibitemOpen
  \bibfield  {author} {\bibinfo {author} {\bibfnamefont {J.~T.}\ \bibnamefont
  {Lizier}}, \bibinfo {author} {\bibfnamefont {M.}~\bibnamefont {Prokopenko}},
  \ and\ \bibinfo {author} {\bibfnamefont {A.~Y.}\ \bibnamefont {Zomaya}},\
  }\bibfield  {title} {\enquote {\bibinfo {title} {Information modification and
  particle collisions in distributed computation},}\ }\href {\doibase
  10.1063/1.3486801} {\bibfield  {journal} {\bibinfo  {journal} {Chaos}\
  }\textbf {\bibinfo {volume} {20}},\ \bibinfo {pages} {037109} (\bibinfo
  {year} {2010})}\BibitemShut {NoStop}%
\bibitem [{\citenamefont {James}, \citenamefont {Ellison},\ and\ \citenamefont
  {Crutchfield}(2011)}]{james2011anatomy}%
  \BibitemOpen
  \bibfield  {author} {\bibinfo {author} {\bibfnamefont {R.~G.}\ \bibnamefont
  {James}}, \bibinfo {author} {\bibfnamefont {C.~J.}\ \bibnamefont {Ellison}},
  \ and\ \bibinfo {author} {\bibfnamefont {J.~P.}\ \bibnamefont
  {Crutchfield}},\ }\bibfield  {title} {\enquote {\bibinfo {title} {Anatomy of
  a bit: Information in a time series observation},}\ }\href {\doibase
  10.1063/1.3637494} {\bibfield  {journal} {\bibinfo  {journal} {Chaos}\
  }\textbf {\bibinfo {volume} {21}},\ \bibinfo {pages} {037109} (\bibinfo
  {year} {2011})}\BibitemShut {NoStop}%
\bibitem [{\citenamefont {Hellman}\ and\ \citenamefont
  {Raviv}(1970)}]{hellman1970probability}%
  \BibitemOpen
  \bibfield  {author} {\bibinfo {author} {\bibfnamefont {M.}~\bibnamefont
  {Hellman}}\ and\ \bibinfo {author} {\bibfnamefont {J.}~\bibnamefont
  {Raviv}},\ }\bibfield  {title} {\enquote {\bibinfo {title} {Probability of
  error, equivocation, and the {Chernoff} bound},}\ }\href {\doibase
  10.1109/TIT.1970.1054466} {\bibfield  {journal} {\bibinfo  {journal} {IEEE
  Transactions on Information Theory}\ }\textbf {\bibinfo {volume} {16}},\
  \bibinfo {pages} {368--372} (\bibinfo {year} {1970})}\BibitemShut {NoStop}%
\bibitem [{\citenamefont {Feder}\ and\ \citenamefont
  {Merhav}(1994)}]{feder1994relations}%
  \BibitemOpen
  \bibfield  {author} {\bibinfo {author} {\bibfnamefont {M.}~\bibnamefont
  {Feder}}\ and\ \bibinfo {author} {\bibfnamefont {N.}~\bibnamefont {Merhav}},\
  }\bibfield  {title} {\enquote {\bibinfo {title} {Relations between entropy
  and error probability},}\ }\href {\doibase 10.1109/18.272494} {\bibfield
  {journal} {\bibinfo  {journal} {IEEE Transactions on Information Theory}\
  }\textbf {\bibinfo {volume} {40}},\ \bibinfo {pages} {259--266} (\bibinfo
  {year} {1994})}\BibitemShut {NoStop}%
\bibitem [{\citenamefont {Cohen}\ \emph {et~al.}(2020)\citenamefont {Cohen},
  \citenamefont {Sasai}, \citenamefont {Tsuchiya},\ and\ \citenamefont
  {Oizumi}}]{cohen2020general}%
  \BibitemOpen
  \bibfield  {author} {\bibinfo {author} {\bibfnamefont {D.}~\bibnamefont
  {Cohen}}, \bibinfo {author} {\bibfnamefont {S.}~\bibnamefont {Sasai}},
  \bibinfo {author} {\bibfnamefont {N.}~\bibnamefont {Tsuchiya}}, \ and\
  \bibinfo {author} {\bibfnamefont {M.}~\bibnamefont {Oizumi}},\ }\bibfield
  {title} {\enquote {\bibinfo {title} {A general spectral decomposition of
  causal influences applied to integrated information},}\ }\href {\doibase
  10.1016/j.jneumeth.2019.108443} {\bibfield  {journal} {\bibinfo  {journal}
  {Journal of Neuroscience Methods}\ }\textbf {\bibinfo {volume} {330}},\
  \bibinfo {pages} {108443} (\bibinfo {year} {2020})}\BibitemShut {NoStop}%
\bibitem [{\citenamefont {Langer}\ and\ \citenamefont {Ay}(2020)}]{Langer2020}%
  \BibitemOpen
  \bibfield  {author} {\bibinfo {author} {\bibfnamefont {C.}~\bibnamefont
  {Langer}}\ and\ \bibinfo {author} {\bibfnamefont {N.}~\bibnamefont {Ay}},\
  }\bibfield  {title} {\enquote {\bibinfo {title} {Complexity as causal
  information integration},}\ }\href {\doibase 10.3390/e22101107} {\bibfield
  {journal} {\bibinfo  {journal} {Entropy}\ }\textbf {\bibinfo {volume} {22}},\
  \bibinfo {pages} {1107} (\bibinfo {year} {2020})}\BibitemShut {NoStop}%
\bibitem [{\citenamefont {Kuramoto}(1984)}]{Kuramoto1984}%
  \BibitemOpen
  \bibfield  {author} {\bibinfo {author} {\bibfnamefont {Y.}~\bibnamefont
  {Kuramoto}},\ }\href@noop {} {\emph {\bibinfo {title} {{Chemical
  Oscillations, Waves and Turbulence}}}}\ (\bibinfo  {publisher} {Dover
  Publications},\ \bibinfo {year} {1984})\ p.\ \bibinfo {pages}
  {164}\BibitemShut {NoStop}%
\bibitem [{\citenamefont {Panaggio}\ and\ \citenamefont
  {Abrams}(2015)}]{Panaggio2015}%
  \BibitemOpen
  \bibfield  {author} {\bibinfo {author} {\bibfnamefont {M.~J.}\ \bibnamefont
  {Panaggio}}\ and\ \bibinfo {author} {\bibfnamefont {D.~M.}\ \bibnamefont
  {Abrams}},\ }\bibfield  {title} {\enquote {\bibinfo {title} {Chimera states:
  {Coexistence} of coherence and incoherence in networks of coupled
  oscillators},}\ }\href {\doibase 10.1088/0951-7715/28/3/R67} {\bibfield
  {journal} {\bibinfo  {journal} {Nonlinearity}\ }\textbf {\bibinfo {volume}
  {28}},\ \bibinfo {pages} {R67--R87} (\bibinfo {year} {2015})},\ \Eprint
  {http://arxiv.org/abs/1403.6204} {arXiv:1403.6204} \BibitemShut {NoStop}%
\bibitem [{\citenamefont {Shanahan}(2010)}]{Shanahan2010}%
  \BibitemOpen
  \bibfield  {author} {\bibinfo {author} {\bibfnamefont {M.}~\bibnamefont
  {Shanahan}},\ }\bibfield  {title} {\enquote {\bibinfo {title} {Metastable
  chimera states in community-structured oscillator networks},}\ }\href
  {\doibase 10.1063/1.3305451} {\bibfield  {journal} {\bibinfo  {journal}
  {Chaos}\ }\textbf {\bibinfo {volume} {20}},\ \bibinfo {pages} {013108}
  (\bibinfo {year} {2010})},\ \Eprint {http://arxiv.org/abs/0908.3881}
  {arXiv:0908.3881} \BibitemShut {NoStop}%
\bibitem [{\citenamefont {Cabral}\ \emph {et~al.}(2011)\citenamefont {Cabral},
  \citenamefont {Hugues}, \citenamefont {Sporns},\ and\ \citenamefont
  {Deco}}]{Cabral2011}%
  \BibitemOpen
  \bibfield  {author} {\bibinfo {author} {\bibfnamefont {J.}~\bibnamefont
  {Cabral}}, \bibinfo {author} {\bibfnamefont {E.}~\bibnamefont {Hugues}},
  \bibinfo {author} {\bibfnamefont {O.}~\bibnamefont {Sporns}}, \ and\ \bibinfo
  {author} {\bibfnamefont {G.}~\bibnamefont {Deco}},\ }\bibfield  {title}
  {\enquote {\bibinfo {title} {{Role of local network oscillations in
  resting-state functional connectivity}},}\ }\href {\doibase
  10.1016/j.neuroimage.2011.04.010} {\bibfield  {journal} {\bibinfo  {journal}
  {NeuroImage}\ }\textbf {\bibinfo {volume} {57}},\ \bibinfo {pages} {130--9}
  (\bibinfo {year} {2011})}\BibitemShut {NoStop}%
\bibitem [{\citenamefont {Schartner}\ \emph {et~al.}(2015)\citenamefont
  {Schartner}, \citenamefont {Seth}, \citenamefont {Noirhomme}, \citenamefont
  {Boly}, \citenamefont {Bruno}, \citenamefont {Laureys},\ and\ \citenamefont
  {Barrett}}]{Schartner2015}%
  \BibitemOpen
  \bibfield  {author} {\bibinfo {author} {\bibfnamefont {M.}~\bibnamefont
  {Schartner}}, \bibinfo {author} {\bibfnamefont {A.~K.}\ \bibnamefont {Seth}},
  \bibinfo {author} {\bibfnamefont {Q.}~\bibnamefont {Noirhomme}}, \bibinfo
  {author} {\bibfnamefont {M.}~\bibnamefont {Boly}}, \bibinfo {author}
  {\bibfnamefont {M.-A.}\ \bibnamefont {Bruno}}, \bibinfo {author}
  {\bibfnamefont {S.}~\bibnamefont {Laureys}}, \ and\ \bibinfo {author}
  {\bibfnamefont {A.~B.}\ \bibnamefont {Barrett}},\ }\bibfield  {title}
  {\enquote {\bibinfo {title} {Complexity of multi-dimensional spontaneous
  {EEG} decreases during propofol induced general anaesthesia},}\ }\href
  {\doibase 10.1371/journal.pone.0133532} {\bibfield  {journal} {\bibinfo
  {journal} {PLoS ONE}\ }\textbf {\bibinfo {volume} {10}},\ \bibinfo {pages}
  {e0133532} (\bibinfo {year} {2015})}\BibitemShut {NoStop}%
\bibitem [{Note1()}]{Note1}%
  \BibitemOpen
  \bibinfo {note} {This can be verified by performing a random time-shuffle on
  the time series, which leaves $\lambda $ and $H_c$ unaltered as they do not
  explicitly depend on time correlations, but makes $\protect \ensuremath
  {\protect \ensuremath {\Phi }\protect \xspace ^{\protect \mathrm
  {R}}}\protect \xspace $ shrink to zero.}\BibitemShut {Stop}%
\bibitem [{\citenamefont {Mitchell}(1996)}]{mitchell1996computation}%
  \BibitemOpen
  \bibfield  {author} {\bibinfo {author} {\bibfnamefont {M.}~\bibnamefont
  {Mitchell}},\ }\bibfield  {title} {\enquote {\bibinfo {title} {Computation in
  cellular automata: A selected review},}\ }in\ \href {\doibase
  10.1002/3527602968.ch4} {\emph {\bibinfo {booktitle} {Non-Standard
  Computation}}},\ \bibinfo {editor} {edited by\ \bibinfo {editor}
  {\bibfnamefont {T.}~\bibnamefont {Gram{\ss}}}, \bibinfo {editor}
  {\bibfnamefont {S.}~\bibnamefont {Bornholdt}}, \bibinfo {editor}
  {\bibfnamefont {M.}~\bibnamefont {Gro{\ss}}}, \bibinfo {editor}
  {\bibfnamefont {M.}~\bibnamefont {Mitchell}}, \ and\ \bibinfo {editor}
  {\bibfnamefont {T.}~\bibnamefont {Pellizzari}}}\ (\bibinfo  {publisher}
  {Wiley-VCH},\ \bibinfo {year} {1996})\ pp.\ \bibinfo {pages}
  {95--140}\BibitemShut {NoStop}%
\bibitem [{\citenamefont {Rosas}\ \emph {et~al.}(2018)\citenamefont {Rosas},
  \citenamefont {Mediano}, \citenamefont {Ugarte},\ and\ \citenamefont
  {Jensen}}]{Rosas2018}%
  \BibitemOpen
  \bibfield  {author} {\bibinfo {author} {\bibfnamefont {F.}~\bibnamefont
  {Rosas}}, \bibinfo {author} {\bibfnamefont {P.~A.}\ \bibnamefont {Mediano}},
  \bibinfo {author} {\bibfnamefont {M.}~\bibnamefont {Ugarte}}, \ and\ \bibinfo
  {author} {\bibfnamefont {H.~J.}\ \bibnamefont {Jensen}},\ }\bibfield  {title}
  {\enquote {\bibinfo {title} {An information-theoretic approach to
  self-organisation: Emergence of complex interdependencies in coupled
  dynamical systems},}\ }\href {\doibase 10.3390/e20100793} {\bibfield
  {journal} {\bibinfo  {journal} {Entropy}\ }\textbf {\bibinfo {volume} {20}},\
  \bibinfo {pages} {793} (\bibinfo {year} {2018})}\BibitemShut {NoStop}%
\bibitem [{\citenamefont {Culik~II}\ and\ \citenamefont
  {Yu}(1988)}]{Culik1988}%
  \BibitemOpen
  \bibfield  {author} {\bibinfo {author} {\bibfnamefont {K.}~\bibnamefont
  {Culik~II}}\ and\ \bibinfo {author} {\bibfnamefont {S.}~\bibnamefont {Yu}},\
  }\bibfield  {title} {\enquote {\bibinfo {title} {Undecidability of {CA}
  classification schemes},}\ }\href@noop {} {\bibfield  {journal} {\bibinfo
  {journal} {Complex Systems}\ }\textbf {\bibinfo {volume} {2}},\ \bibinfo
  {pages} {177--190} (\bibinfo {year} {1988})}\BibitemShut {NoStop}%
\bibitem [{\citenamefont {Martinez}, \citenamefont {Seck-Tuoh-Mora},\ and\
  \citenamefont {Zenil}(2013)}]{Martinez2013}%
  \BibitemOpen
  \bibfield  {author} {\bibinfo {author} {\bibfnamefont {G.~J.}\ \bibnamefont
  {Martinez}}, \bibinfo {author} {\bibfnamefont {J.~C.}\ \bibnamefont
  {Seck-Tuoh-Mora}}, \ and\ \bibinfo {author} {\bibfnamefont {H.}~\bibnamefont
  {Zenil}},\ }\bibfield  {title} {\enquote {\bibinfo {title} {Computation and
  universality: {Class IV} versus class {III} cellular automata},}\ }\href@noop
  {} {\bibfield  {journal} {\bibinfo  {journal} {Journal of Cellular Automata}\
  }\textbf {\bibinfo {volume} {7}},\ \bibinfo {pages} {393--430} (\bibinfo
  {year} {2013})},\ \Eprint {http://arxiv.org/abs/1304.1242} {arXiv:1304.1242}
  \BibitemShut {NoStop}%
\bibitem [{\citenamefont {Scheffer}\ \emph {et~al.}(2009)\citenamefont
  {Scheffer}, \citenamefont {Bascompte}, \citenamefont {Brock}, \citenamefont
  {Brovkin}, \citenamefont {Carpenter}, \citenamefont {Dakos}, \citenamefont
  {Held}, \citenamefont {Van~Nes}, \citenamefont {Rietkerk},\ and\
  \citenamefont {Sugihara}}]{Scheffer2009}%
  \BibitemOpen
  \bibfield  {author} {\bibinfo {author} {\bibfnamefont {M.}~\bibnamefont
  {Scheffer}}, \bibinfo {author} {\bibfnamefont {J.}~\bibnamefont {Bascompte}},
  \bibinfo {author} {\bibfnamefont {W.~A.}\ \bibnamefont {Brock}}, \bibinfo
  {author} {\bibfnamefont {V.}~\bibnamefont {Brovkin}}, \bibinfo {author}
  {\bibfnamefont {S.~R.}\ \bibnamefont {Carpenter}}, \bibinfo {author}
  {\bibfnamefont {V.}~\bibnamefont {Dakos}}, \bibinfo {author} {\bibfnamefont
  {H.}~\bibnamefont {Held}}, \bibinfo {author} {\bibfnamefont {E.~H.}\
  \bibnamefont {Van~Nes}}, \bibinfo {author} {\bibfnamefont {M.}~\bibnamefont
  {Rietkerk}}, \ and\ \bibinfo {author} {\bibfnamefont {G.}~\bibnamefont
  {Sugihara}},\ }\bibfield  {title} {\enquote {\bibinfo {title} {Early-warning
  signals for critical transitions},}\ }\href {\doibase 10.1038/nature08227}
  {\bibfield  {journal} {\bibinfo  {journal} {Nature}\ }\textbf {\bibinfo
  {volume} {461}},\ \bibinfo {pages} {53} (\bibinfo {year} {2009})}\BibitemShut
  {NoStop}%
\bibitem [{Note2()}]{Note2}%
  \BibitemOpen
  \bibinfo {note} {Indeed, although some pointwise redundancy functions
  exist,\cite {Finn2018} the relatively much simpler redundancy in Eq.~\protect
  \textup {\hbox {\mathsurround \z@ \protect \normalfont (\ignorespaces \ref
  {eq:phiR}\unskip \@@italiccorr )}} is not naturally translated into a
  pointwise setting, and thus we restrict the pointwise analyses to standard
  integrated information only.}\BibitemShut {Stop}%
\bibitem [{\citenamefont {Kelso}(1995)}]{Kelso1995}%
  \BibitemOpen
  \bibfield  {author} {\bibinfo {author} {\bibfnamefont {J.~S.}\ \bibnamefont
  {Kelso}},\ }\href@noop {} {\emph {\bibinfo {title} {Dynamic Patterns: {The}
  Self-organization of Brain and Behavior}}}\ (\bibinfo  {publisher} {MIT
  Press},\ \bibinfo {year} {1995})\BibitemShut {NoStop}%
\bibitem [{\citenamefont {Rabinovich}\ \emph {et~al.}(2008)\citenamefont
  {Rabinovich}, \citenamefont {Huerta}, \citenamefont {Varona},\ and\
  \citenamefont {Afraimovich}}]{rabinovich2008transient}%
  \BibitemOpen
  \bibfield  {author} {\bibinfo {author} {\bibfnamefont {M.~I.}\ \bibnamefont
  {Rabinovich}}, \bibinfo {author} {\bibfnamefont {R.}~\bibnamefont {Huerta}},
  \bibinfo {author} {\bibfnamefont {P.}~\bibnamefont {Varona}}, \ and\ \bibinfo
  {author} {\bibfnamefont {V.~S.}\ \bibnamefont {Afraimovich}},\ }\bibfield
  {title} {\enquote {\bibinfo {title} {Transient cognitive dynamics,
  metastability, and decision making},}\ }\href {\doibase
  10.1371/journal.pcbi.1000072} {\bibfield  {journal} {\bibinfo  {journal}
  {PLoS Computational Biology}\ }\textbf {\bibinfo {volume} {4}},\ \bibinfo
  {pages} {e1000072} (\bibinfo {year} {2008})}\BibitemShut {NoStop}%
\bibitem [{\citenamefont {Beggs}\ and\ \citenamefont
  {Timme}(2012)}]{beggs2012being}%
  \BibitemOpen
  \bibfield  {author} {\bibinfo {author} {\bibfnamefont {J.~M.}\ \bibnamefont
  {Beggs}}\ and\ \bibinfo {author} {\bibfnamefont {N.}~\bibnamefont {Timme}},\
  }\bibfield  {title} {\enquote {\bibinfo {title} {Being critical of
  criticality in the brain},}\ }\href {\doibase 10.3389/fphys.2012.00163}
  {\bibfield  {journal} {\bibinfo  {journal} {Frontiers in Physiology}\
  }\textbf {\bibinfo {volume} {3}},\ \bibinfo {pages} {163} (\bibinfo {year}
  {2012})}\BibitemShut {NoStop}%
\bibitem [{\citenamefont {Pruessner}(2012)}]{pruessner2012self}%
  \BibitemOpen
  \bibfield  {author} {\bibinfo {author} {\bibfnamefont {G.}~\bibnamefont
  {Pruessner}},\ }\href@noop {} {\emph {\bibinfo {title} {Self-organised
  Criticality: Theory, Models and Characterisation}}}\ (\bibinfo  {publisher}
  {Cambridge University Press},\ \bibinfo {year} {2012})\BibitemShut {NoStop}%
\bibitem [{\citenamefont {Tononi}, \citenamefont {Sporns},\ and\ \citenamefont
  {Edelman}(1994)}]{Tononi1994}%
  \BibitemOpen
  \bibfield  {author} {\bibinfo {author} {\bibfnamefont {G.}~\bibnamefont
  {Tononi}}, \bibinfo {author} {\bibfnamefont {O.}~\bibnamefont {Sporns}}, \
  and\ \bibinfo {author} {\bibfnamefont {G.~M.}\ \bibnamefont {Edelman}},\
  }\bibfield  {title} {\enquote {\bibinfo {title} {A measure for brain
  complexity: {Relating} functional segregation and integration in the nervous
  system},}\ }\href {\doibase 10.1073/pnas.91.11.5033} {\bibfield  {journal}
  {\bibinfo  {journal} {Proceedings of the National Academy of Sciences}\
  }\textbf {\bibinfo {volume} {91}},\ \bibinfo {pages} {5033--7} (\bibinfo
  {year} {1994})}\BibitemShut {NoStop}%
\bibitem [{\citenamefont {Rosas}\ \emph {et~al.}(2019)\citenamefont {Rosas},
  \citenamefont {Mediano}, \citenamefont {Gastpar},\ and\ \citenamefont
  {Jensen}}]{Rosas2019}%
  \BibitemOpen
  \bibfield  {author} {\bibinfo {author} {\bibfnamefont {F.~E.}\ \bibnamefont
  {Rosas}}, \bibinfo {author} {\bibfnamefont {P.~A.}\ \bibnamefont {Mediano}},
  \bibinfo {author} {\bibfnamefont {M.}~\bibnamefont {Gastpar}}, \ and\
  \bibinfo {author} {\bibfnamefont {H.~J.}\ \bibnamefont {Jensen}},\ }\bibfield
   {title} {\enquote {\bibinfo {title} {Quantifying high-order
  interdependencies via multivariate extensions of the mutual information},}\
  }\href {\doibase 10.1103/PhysRevE.100.032305} {\bibfield  {journal} {\bibinfo
   {journal} {Physical Review E}\ }\textbf {\bibinfo {volume} {100}},\ \bibinfo
  {pages} {032305} (\bibinfo {year} {2019})}\BibitemShut {NoStop}%
\bibitem [{\citenamefont {Rosas}\ \emph {et~al.}(2020)\citenamefont {Rosas},
  \citenamefont {Mediano}, \citenamefont {Jensen}, \citenamefont {Seth},
  \citenamefont {Barrett}, \citenamefont {Carhart-Harris},\ and\ \citenamefont
  {Bor}}]{rosas2020reconciling}%
  \BibitemOpen
  \bibfield  {author} {\bibinfo {author} {\bibfnamefont {F.~E.}\ \bibnamefont
  {Rosas}}, \bibinfo {author} {\bibfnamefont {P.~A.}\ \bibnamefont {Mediano}},
  \bibinfo {author} {\bibfnamefont {H.~J.}\ \bibnamefont {Jensen}}, \bibinfo
  {author} {\bibfnamefont {A.~K.}\ \bibnamefont {Seth}}, \bibinfo {author}
  {\bibfnamefont {A.~B.}\ \bibnamefont {Barrett}}, \bibinfo {author}
  {\bibfnamefont {R.~L.}\ \bibnamefont {Carhart-Harris}}, \ and\ \bibinfo
  {author} {\bibfnamefont {D.}~\bibnamefont {Bor}},\ }\bibfield  {title}
  {\enquote {\bibinfo {title} {Reconciling emergences: {An}
  information-theoretic approach to identify causal emergence in multivariate
  data},}\ }\href {\doibase 10.1371/journal.pcbi.1008289} {\bibfield  {journal}
  {\bibinfo  {journal} {PLoS Computational Biology}\ }\textbf {\bibinfo
  {volume} {16}},\ \bibinfo {pages} {1--22} (\bibinfo {year}
  {2020})}\BibitemShut {NoStop}%
\bibitem [{\citenamefont {Tononi}, \citenamefont {Edelman},\ and\ \citenamefont
  {Sporns}(1998)}]{Tononi1998}%
  \BibitemOpen
  \bibfield  {author} {\bibinfo {author} {\bibfnamefont {G.}~\bibnamefont
  {Tononi}}, \bibinfo {author} {\bibfnamefont {G.~M.}\ \bibnamefont {Edelman}},
  \ and\ \bibinfo {author} {\bibfnamefont {O.}~\bibnamefont {Sporns}},\
  }\bibfield  {title} {\enquote {\bibinfo {title} {Complexity and coherency:
  {Integrating} information in the brain},}\ }\href {\doibase
  10.1016/S1364-6613(98)01259-5} {\bibfield  {journal} {\bibinfo  {journal}
  {Trends in Cognitive Sciences}\ }\textbf {\bibinfo {volume} {2}},\ \bibinfo
  {pages} {474--484} (\bibinfo {year} {1998})}\BibitemShut {NoStop}%
\bibitem [{\citenamefont {Sporns}, \citenamefont {Tononi},\ and\ \citenamefont
  {Edelman}(2000)}]{Sporns2000}%
  \BibitemOpen
  \bibfield  {author} {\bibinfo {author} {\bibfnamefont {O.}~\bibnamefont
  {Sporns}}, \bibinfo {author} {\bibfnamefont {G.}~\bibnamefont {Tononi}}, \
  and\ \bibinfo {author} {\bibfnamefont {G.~M.}\ \bibnamefont {Edelman}},\
  }\bibfield  {title} {\enquote {\bibinfo {title} {Connectivity and complexity:
  The relationship between neuroanatomy and brain dynamics},}\ }\href {\doibase
  10.1016/s0893-6080(00)00053-8} {\bibfield  {journal} {\bibinfo  {journal}
  {Neural Networks}\ }\textbf {\bibinfo {volume} {13}},\ \bibinfo {pages}
  {909--922} (\bibinfo {year} {2000})}\BibitemShut {NoStop}%
\bibitem [{\citenamefont {Ay}(2015)}]{Ay2015}%
  \BibitemOpen
  \bibfield  {author} {\bibinfo {author} {\bibfnamefont {N.}~\bibnamefont
  {Ay}},\ }\bibfield  {title} {\enquote {\bibinfo {title} {Information geometry
  on complexity and stochastic interaction},}\ }\href {\doibase
  10.3390/e17042432} {\bibfield  {journal} {\bibinfo  {journal} {Entropy}\
  }\textbf {\bibinfo {volume} {17}},\ \bibinfo {pages} {2432--2458} (\bibinfo
  {year} {2015})}\BibitemShut {NoStop}%
\bibitem [{\citenamefont {Oizumi}, \citenamefont {Tsuchiya},\ and\
  \citenamefont {Amari}(2016)}]{Oizumi2016}%
  \BibitemOpen
  \bibfield  {author} {\bibinfo {author} {\bibfnamefont {M.}~\bibnamefont
  {Oizumi}}, \bibinfo {author} {\bibfnamefont {N.}~\bibnamefont {Tsuchiya}}, \
  and\ \bibinfo {author} {\bibfnamefont {S.-i.}\ \bibnamefont {Amari}},\
  }\bibfield  {title} {\enquote {\bibinfo {title} {Unified framework for
  information integration based on information geometry},}\ }\href {\doibase
  10.1073/pnas.1603583113} {\bibfield  {journal} {\bibinfo  {journal}
  {Proceedings of the National Academy of Sciences}\ }\textbf {\bibinfo
  {volume} {113}},\ \bibinfo {pages} {14817--14822} (\bibinfo {year}
  {2016})}\BibitemShut {NoStop}%
\bibitem [{\citenamefont {Tagliazucchi}(2017)}]{Tagliazucchi2017}%
  \BibitemOpen
  \bibfield  {author} {\bibinfo {author} {\bibfnamefont {E.}~\bibnamefont
  {Tagliazucchi}},\ }\bibfield  {title} {\enquote {\bibinfo {title} {The
  signatures of conscious access and its phenomenology are consistent with
  large-scale brain communication at criticality},}\ }\href {\doibase
  10.1016/j.concog.2017.08.008} {\bibfield  {journal} {\bibinfo  {journal}
  {Consciousness and Cognition}\ }\textbf {\bibinfo {volume} {55}},\ \bibinfo
  {pages} {136--147} (\bibinfo {year} {2017})}\BibitemShut {NoStop}%
\bibitem [{\citenamefont {Mediano}\ and\ \citenamefont
  {Shanahan}(2017)}]{Mediano2017}%
  \BibitemOpen
  \bibfield  {author} {\bibinfo {author} {\bibfnamefont {P.}~\bibnamefont
  {Mediano}}\ and\ \bibinfo {author} {\bibfnamefont {M.}~\bibnamefont
  {Shanahan}},\ }\bibfield  {title} {\enquote {\bibinfo {title} {Balanced
  information storage and transfer in modular spiking neural networks},}\
  }\href@noop {} {\  (\bibinfo {year} {2017})},\ \Eprint
  {http://arxiv.org/abs/1708.04392} {arXiv:1708.04392} \BibitemShut {NoStop}%
\bibitem [{\citenamefont {Aguilera}(2019)}]{Aguilera2019}%
  \BibitemOpen
  \bibfield  {author} {\bibinfo {author} {\bibfnamefont {M.}~\bibnamefont
  {Aguilera}},\ }\bibfield  {title} {\enquote {\bibinfo {title} {Scaling
  behaviour and critical phase transitions in integrated information theory},}\
  }\href {\doibase 10.3390/e21121198} {\bibfield  {journal} {\bibinfo
  {journal} {Entropy}\ }\textbf {\bibinfo {volume} {21}},\ \bibinfo {pages}
  {1198} (\bibinfo {year} {2019})}\BibitemShut {NoStop}%
\bibitem [{\citenamefont {Parrondo}, \citenamefont {Horowitz},\ and\
  \citenamefont {Sagawa}(2015)}]{parrondo2015thermodynamics}%
  \BibitemOpen
  \bibfield  {author} {\bibinfo {author} {\bibfnamefont {J.~M.}\ \bibnamefont
  {Parrondo}}, \bibinfo {author} {\bibfnamefont {J.~M.}\ \bibnamefont
  {Horowitz}}, \ and\ \bibinfo {author} {\bibfnamefont {T.}~\bibnamefont
  {Sagawa}},\ }\bibfield  {title} {\enquote {\bibinfo {title} {Thermodynamics
  of information},}\ }\href@noop {} {\bibfield  {journal} {\bibinfo  {journal}
  {Nature physics}\ }\textbf {\bibinfo {volume} {11}},\ \bibinfo {pages}
  {131--139} (\bibinfo {year} {2015})}\BibitemShut {NoStop}%
\bibitem [{\citenamefont {Prokopenko}\ \emph {et~al.}(2011)\citenamefont
  {Prokopenko}, \citenamefont {Lizier}, \citenamefont {Obst},\ and\
  \citenamefont {Wang}}]{Prokopenko2011}%
  \BibitemOpen
  \bibfield  {author} {\bibinfo {author} {\bibfnamefont {M.}~\bibnamefont
  {Prokopenko}}, \bibinfo {author} {\bibfnamefont {J.~T.}\ \bibnamefont
  {Lizier}}, \bibinfo {author} {\bibfnamefont {O.}~\bibnamefont {Obst}}, \ and\
  \bibinfo {author} {\bibfnamefont {X.~R.}\ \bibnamefont {Wang}},\ }\bibfield
  {title} {\enquote {\bibinfo {title} {Relating {Fisher} information to order
  parameters},}\ }\href {\doibase 10.1103/PhysRevE.84.041116} {\bibfield
  {journal} {\bibinfo  {journal} {Physical Review E}\ }\textbf {\bibinfo
  {volume} {84}},\ \bibinfo {pages} {041116} (\bibinfo {year}
  {2011})}\BibitemShut {NoStop}%
\bibitem [{\citenamefont {Crutchfield}\ and\ \citenamefont
  {Feldman}(2003)}]{Crutchfield2003}%
  \BibitemOpen
  \bibfield  {author} {\bibinfo {author} {\bibfnamefont {J.~P.}\ \bibnamefont
  {Crutchfield}}\ and\ \bibinfo {author} {\bibfnamefont {D.~P.}\ \bibnamefont
  {Feldman}},\ }\bibfield  {title} {\enquote {\bibinfo {title} {Regularities
  unseen, randomness observed: {Levels} of entropy convergence},}\ }\href
  {\doibase 10.1063/1.1530990} {\bibfield  {journal} {\bibinfo  {journal}
  {Chaos}\ }\textbf {\bibinfo {volume} {13}},\ \bibinfo {pages} {25--54}
  (\bibinfo {year} {2003})}\BibitemShut {NoStop}%
\bibitem [{\citenamefont {Schreiber}(2000)}]{Schreiber2000}%
  \BibitemOpen
  \bibfield  {author} {\bibinfo {author} {\bibfnamefont {T.}~\bibnamefont
  {Schreiber}},\ }\bibfield  {title} {\enquote {\bibinfo {title} {Measuring
  information transfer},}\ }\href {\doibase 10.1103/PhysRevLett.85.461}
  {\bibfield  {journal} {\bibinfo  {journal} {Physical Review Letters}\
  }\textbf {\bibinfo {volume} {85}},\ \bibinfo {pages} {461--464} (\bibinfo
  {year} {2000})}\BibitemShut {NoStop}%
\bibitem [{\citenamefont {Finn}\ and\ \citenamefont {Lizier}(2018)}]{Finn2018}%
  \BibitemOpen
  \bibfield  {author} {\bibinfo {author} {\bibfnamefont {C.}~\bibnamefont
  {Finn}}\ and\ \bibinfo {author} {\bibfnamefont {J.~T.}\ \bibnamefont
  {Lizier}},\ }\bibfield  {title} {\enquote {\bibinfo {title} {Pointwise
  partial information decomposition using the specificity and ambiguity
  lattices},}\ }\href {\doibase 10.3390/e20040297} {\bibfield  {journal}
  {\bibinfo  {journal} {Entropy}\ }\textbf {\bibinfo {volume} {20}},\ \bibinfo
  {pages} {297} (\bibinfo {year} {2018})}\BibitemShut {NoStop}%
\end{thebibliography}%

\end{document}